\documentclass[sigconf,nonacm,screen,pdfa]{acmart}
\pdfoutput=1
\usepackage[utf8]{inputenc}
\usepackage{tcolorbox}
\usepackage{graphicx}
\usepackage{multirow,makecell}
\usepackage[ruled,vlined,linesnumbered]{algorithm2e}
\usepackage{subcaption}
\usepackage{booktabs}
\usepackage{pifont}
\usepackage{lipsum}
\usepackage{makecell}
\usepackage[para,online,flushleft]{threeparttable}
\usepackage{mathtools}
\usepackage{listings}
\usepackage{wrapfig}
\usepackage{balance}
\usepackage{amsmath}
\usepackage{adjustbox}
\usepackage{tabularx}
\usepackage[capitalise,noabbrev]{cleveref}
\usepackage{fancybox}
\usepackage[framemethod=tikz]{mdframed}
\usepackage{enumitem}
\usepackage{scalerel}
\usepackage{microtype}

\usepackage{hyperref}
\hypersetup{
    colorlinks=true,
    linkcolor=blue,
    filecolor=magenta,      
    urlcolor=cyan,
}

\title{Hydro: Adaptive Query Processing of ML Queries}

\author{Gaurav Tarlok Kakkar}
\authornotemark[1]
\affiliation{Georgia Institute of Technology}
\email{gkakkar7@gatech.edu}

\author{Jiashen Cao}
\authornote{Equal contribution (order determined by dice roll).}
\affiliation{Georgia Institute of Technology}
\email{jiashenc@gatech.edu}

\author{Aubhro Sengupta}
\affiliation{Georgia Institute of Technology}
\email{aubhros@gatech.edu}

\author{Joy Arulraj}
\affiliation{Georgia Institute of Technology}
\email{arulraj@gatech.edu}

\author{Hyesoon Kim}
\affiliation{Georgia Institute of Technology}
\email{hyesoon@cc.gatech.edu}

\definecolor{linkcolor}{HTML}{647382}
\definecolor{citecolor}{HTML}{647382} %
\definecolor{urlcolor}{rgb}{0.4,0.2,0.2}
\definecolor{sqlcolor}{HTML}{965d67}
\definecolor{smtcolor}{HTML}{5d968c}
\definecolor{webblue}{rgb}{0,0,.7}
\definecolor{webgreen}{rgb}{0,.5,0}
\definecolor{webbrown}{rgb}{.6,0,0}

\usetikzlibrary{shadows}
\tikzset{every shadow/.style={opacity=1}}
\newmdenv[shadow=false,shadowcolor=black,shadowsize=0pt,linewidth=1pt,skipabove=2pt]{highlightbox}

\makeatletter
\def\@ACM@checkaffil{
    \if@ACM@instpresent\else
    \ClassWarningNoLine{\@classname}{No institution present for an affiliation}%
    \fi
    \if@ACM@citypresent\else
    \ClassWarningNoLine{\@classname}{No city present for an affiliation}%
    \fi
    \if@ACM@countrypresent\else
        \ClassWarningNoLine{\@classname}{No country present for an affiliation}%
    \fi
}
\makeatother

\newcommand{\probP}{\text{I\kern-0.15em P}}

\newcommand{\sys}{\textsc{Hydro}\xspace}
\newcommand{\aqp}{\textsc{AQP}\xspace}
\newcommand{\eva}{\textsc{EvaDB}\xspace}

\newcommand{\eddy}{\textsc{Eddy}\xspace}
\newcommand{\laminar}{\textsc{Laminar}\xspace}
\newcommand{\eddypull}{\textsc{Eddy pull}\xspace}
\newcommand{\centralqueue}{\textsc{Central queue}\xspace}
\newcommand{\eddyrouter}{\textsc{Eddy router}\xspace}
\newcommand{\laminarrouter}{\textsc{Laminar router}\xspace}

\newcommand{\eg}{\textit{e.g.,}\xspace}
\newcommand{\ie}{\textit{i.e.,}\xspace}

\newcommand{\etal}{\textit{et al.}\xspace}

\newcommand{\X}{$\times$\xspace}

\newcommand{\udf}{UDF\xspace}
\newcommand{\udfs}{UDFs\xspace}
\newcommand{\sota}{SotA\xspace}
\newcommand{\reuseaware}{\textsc{reuse-aware}\xspace}

\newcommand{\dogbreedudf}{\lstinline[style=SQLStyle]{DogBreedClassifier}\xspace}
\newcommand{\objdetudf}{\lstinline[style=SQLStyle]{ObjectDetector}\xspace}

\newcommand{\PP}[1]{\vspace{4px}\noindent{\bf\textsc{#1.}}\xspace}

\def\Snospace~{\S{}}

\newcommand\BeraMonottfamily{%
  \def\fvm@Scale{0.85}
  \fontfamily{fvm}\selectfont
}

\lstdefinestyle{SQLStyle}{
  language=SQL,
  basicstyle={\footnotesize\ttfamily},
  breaklines=true,
  frame=single, 
  frameround=tttt, 
  backgroundcolor=\color{gray!5}, 
  rulesep=1pt,
  numbers=none,
  keepspaces=true,
  showstringspaces=false,
  captionpos=b,
  aboveskip=5pt,
  belowskip=5pt,
  numberstyle=\tiny\color{gray},  
  stringstyle=\color{webgreen},
  keywordstyle=\color{webblue},
  commentstyle=\color{gray},
  keywords=[2]{LATERAL, UNNEST, APPLY},
  keywordstyle=[2]\color{webblue},
  keywords=[3]{DogBreedClassifier, ObjectDetector, DogColorClassifier, Crop, HardHatDetector, LLM, DogColor, DogBreed},
  keywordstyle=[3]\color{webbrown},
}

\lstdefinestyle{PythonStyle}{
  language=Python,
  basicstyle={\footnotesize\ttfamily},
  breaklines=true,
  frame=single, 
  backgroundcolor=\color{gray!5}, 
  rulesep=1pt,
  numbers=left,
  numberstyle=\tiny\color{black},  
  keepspaces=true,
  showstringspaces=false,
  captionpos=b,
  aboveskip=5pt,
  belowskip=5pt,
  stringstyle=\color{webgreen},
  keywordstyle=\color{webblue},
  commentstyle=\color{gray},
}

\makeatletter
\newcommand*{\rom}[1]{\expandafter\@slowromancap\romannumeral #1@}
\makeatother

\newcommand{\squishitemize}{
 \begin{list}{$\bullet$}
  { \setlength{\itemsep}{0pt}
     \setlength{\parsep}{1pt}
     \setlength{\topsep}{1pt}
     \setlength{\partopsep}{0pt}
     \setlength{\leftmargin}{0.5em}
     \setlength{\labelwidth}{0.5em}
     \setlength{\labelsep}{0.5em} } }

\newcounter{Lcount}
\newcommand{\squishlist}{
    \begin{list}{\arabic{Lcount}. }
   { \usecounter{Lcount}
        \setlength{\itemsep}{0pt}
        \setlength{\parsep}{3pt}
        \setlength{\topsep}{3pt}
        \setlength{\partopsep}{0pt}
        \setlength{\leftmargin}{2em}
        \setlength{\labelwidth}{1.5em}
        \setlength{\labelsep}{0.5em} } }

\newcommand{\squishend}{\end{list}}

\makeatletter
\newcommand{\crefnames}[3]{%
  \@for\next:=#1\do{%
    \expandafter\crefname\expandafter{\next}{#2}{#3}%
  }%
}
\makeatother
\crefnames{part,chapter,section}{\S}{\S\S}
\crefnames{paragraph,subparagraph}{\P}{\P\P}
\crefname{figure}{Fig.}{Figs.}

\newlength{\lineheight}
\begin{document}

\begin{abstract}
Query optimization in relational database management systems (DBMSs) is critical for fast query processing.
The query optimizer relies on precise selectivity and cost estimates to effectively optimize queries prior to execution.
While this strategy is effective for relational DBMSs, it is not sufficient for DBMSs tailored for processing machine learning (ML) queries.
In ML-centric DBMSs, query optimization is challenging for two reasons.
First, the performance bottleneck of the queries shifts to user-defined functions (\udfs) that often wrap around deep learning models, making it difficult to accurately estimate \udf statistics without profiling the query.
This leads to inaccurate statistics and sub-optimal query plans.
Second, the optimal query plan for ML queries is data-dependent, necessitating DBMSs to adapt the query plan on the fly during execution.
So, a static query plan is not sufficient for such queries.

In this paper, we present \sys, an ML-centric DBMS that utilizes adaptive query processing (\aqp) for efficiently processing ML queries.
\sys is designed to quickly evaluate \udf-based query predicates by ensuring optimal predicate evaluation order and improving the scalability of \udf execution.
By integrating \aqp, \sys continuously monitors \udf statistics, routes data to predicates in an optimal order, and dynamically allocates resources for evaluating predicates.
We demonstrate \sys's efficacy through four illustrative use cases, delivering up to $11.52$\X speedup over a baseline system.
\end{abstract}

\settopmatter{printfolios=true}
\maketitle
  
\setcounter{page}{1}

\section{INTRODUCTION}\label{sec:intro}
Modern database systems support ML-powered queries, enabling users to directly execute ML models within the database.
These models are often wrapped within user-defined functions (\udfs).
Since evaluating these functions is computationally expensive, they have become the central focus for query optimization. 
For instance, video database management systems (VDBMSs), which heavily rely on computer vision algorithms, often encounter bottlenecks attributed to \udfs~\cite{xu_eva_2022,cao_figo_2022,romero_optimizing_2022}.

Prior ML-centric DBMSs~\cite{xu_eva_2022,cao_figo_2022,romero_optimizing_2022} have focused on enhancing query plans through \udf statistics derived from sample (canary) data.
However, their static approach to query optimization slows down query processing.
To circumvent this problem, researchers have proposed \aqp frameworks that deliver better query performance over conventional static query optimization~\cite{eddies_avnur_2000, contentaqp_babu_2005}.
An \aqp framework refines the query plan dynamically during execution, offering the potential for superior query plans and improved hardware utilization.
Nevertheless, there has been limited exploration of \aqp's application in scenarios where \udfs are a major performance bottleneck. 
In this paper, we examine new opportunities for leveraging \aqp while processing ML-centric queries.

\PP{Motivation}
Consider a dog owner analyzing surveillance videos to locate their lost pet -- a black-colored great dane.
\begin{lstlisting}[style=SQLStyle, label=lst:intro:example, caption={Query to retrieve frames containing black great dane dogs.}]
SELECT id, bbox FROM video 
JOIN LATERAL UNNEST(ObjectDetector(frame)) AS Object(label, bbox, score) 
WHERE Object.label='dog' 
AND DogBreedClassifier(Crop(frame, bbox)) = 'great dane' 
AND DogColorClassifier(Crop(frame, bbox)) = 'black';
\end{lstlisting}
~\cref{lst:intro:example} presents the corresponding query in a VDBMS for identifying video segments with potential matches. 
The query uses an object detection model to filter frames with dogs.
Subsequently, it applies two predicates: 
one to determine if the dog's bounding box contains a great dane (\lstinline[style=SQLStyle]{DogBreedClassifier(Crop(frame, bbox))='great dane'}), and another to verify if the dog is black-colored (\lstinline[style=SQLStyle]{DogColorClassifier(Crop(frame, bbox))='black'}).
From a query optimization standpoint, the order in which the predicates are evaluated is crucial as these predicates contain computationally expensive \udfs.
Picking the optimal order of predicate evaluation leads to a significant drop in query execution time.
Based on the traditional predicate reordering \cite{hellerstein_practical_1994} technique, the optimal order is determined based on both predicate selectivity and cost.
Typically, the optimizer ranks each predicate using a scoring function, prioritizing inexpensive and highly-selective predicates.
However, this classical approach suffers from two limitations.

\PP{I - Unreliable Statistics}
Prior ML-centric DBMSs either assume that accurate selectivity and cost statistics of the \udfs are already available \cite{xu_eva_2022} or estimate them by running the \udfs over a subset of data \cite{cao_figo_2022,romero_optimizing_2022}. 
However, such estimates can be inaccurate as the \udf evaluation cost may vary throughout the lifespan of a query.
For example, the execution cost of \lstinline[style=SQLStyle]{DogBreedClassifier} is correlated to the bounding box dimension, which varies across data.
Additionally, other system-level optimizations like caching \cite{xu_eva_2022} further reduce the accuracy of the estimates because caching and reusing the outputs of the \udfs directly impact the cost.

Moreover, estimating statistics during query optimization not only increases the query optimization time but also adds complexity to the implementation of the query optimizer.
For instance, in~\cref{lst:intro:example}, to estimate the selectivity of \dogbreedudf, the optimizer must run \objdetudf on the video to extract dogs. 
As these \udfs are computationally expensive, this approach incurs significant optimization overhead.
Furthermore, the input dependency of \dogbreedudf on the output of \objdetudf forces the optimizer to execute its sub-plans to estimate statistics, complicating the query optimizer implementation.

\PP{II - Unscalable \udf Execution}
Secondly, achieving optimal performance with computationally expensive \udfs requires meticulous hardware resource management. 
Given that \udfs often exhibit diverse characteristics, the query execution engine needs to monitor the resource usage of each \udf throughout execution and allocate resources accordingly for optimal performance. 
However, incorporating runtime monitoring and resource adjustment proves challenging within the constraints of a static query optimization scheme.
Additionally, apart from hardware resource management, our findings highlight the significance of effective load balancing between workers during scaling up.
While simple policies like round-robin are adequate for preventing workload imbalance in common scenarios, certain cases demand advanced load-balancing policies that consider data characteristics for workload distribution. 
Consider the example of the \dogbreedudf, where the cost depends on the input dimensions. 
Accounting for this correlation between the cost and the input data is crucial to avoid workload imbalance. 
Our empirical results demonstrate that adopting a data-aware policy results in up to $1.46 \times$ speedup. 
Static query processing cannot accommodate these adaptive, on-the-fly optimizations. 

\PP{Our Approach}
In this paper, we present \sys, an ML-centric DBMS that employs an \aqp scheme, eliminating the need for prior estimates of \udf selectivity and cost. 
\sys improves upon the influential \eddy \aqp framework \cite{eddies_avnur_2000} by tailoring it for ML-centric queries.
\sys optimizes the evaluation order of predicates involving \udfs using the \eddy framework and further enhances the framework by dynamically adjusting \udf hardware resource allocation and optimizing load balancing during execution.

\PP{Contributions}
We make the following contributions:
\squishitemize
\item We propose an ML-centric DBMS, \sys, which adapts the \aqp framework for ML queries. 
Through our proposed ML query-specific routing policies and an adaptive \udfs statistics collection mechanism, \sys ensures an optimal predicate evaluation order for \udf-based query predicates. 
We demonstrate the seamless integration of our framework with other ML query optimizations.
\item We further enhance the \eddy framework by developing an automatic scaling component called \laminar to improve ML query performance through improved hardware utilization. 
We introduce a data-aware routing policy for more effective load balancing during query execution, especially when scaling up.
\item Through experiments spanning four diverse use cases, including three in video analytics and one employing a large language model for analytics, we demonstrate that \sys delivers up to 11.52\X speedup compared to the baseline system. 
\squishend
\section{BACKGROUND}\label{sec:back}
We next discuss the query optimization techniques proposed in \sota DBMSs in~\cref{sec:back:vdbms}.
We then go over the \sota adaptive query processing mechanisms in~\cref{sec:back:aqp}.
\subsection{Query Optimization in VDBMSs}\label{sec:back:vdbms}
VDBMSs often contain novel query optimization techniques tailored for computationally expensive \udfs.~\cite{noscope_kang_2017,blazeit_kang_2019,tahoma_anderson_2019,pp_lu_2018,koudas_video_2020,yang_optimizing_2022,kang_accelerating_2021}.
Xu~\etal~\cite{xu_eva_2022} uses materialized views to store the results of expensive \udfs, thereby accelerating queries in exploratory video analytics settings where queries have overlapping computation.
VIVA~\cite{romero_optimizing_2022} optimizes queries using user-specified relational hints, such as replacement or filtering hints. 
It uses these hints to select the optimal query plan that meets an accuracy constraint.
Both Xu~\etal and VIVA adhere to a static query optimization approach. 
While Xu~\etal assumes the availability of accurate statistics, VIVA estimates them from sample data. 
However, both suffer from poor estimates, leading to suboptimal query plans.
To address this problem, \sys obtains \udfs statistics during query execution and adaptively adjusts the query plan.

ExSample~\cite{exsample_moll_2020} supports distinct object queries and introduces an adaptive sampling algorithm to select frames from video segments likely to contain the object of interest.
FiGO~\cite{cao_figo_2022} harnesses the power of executing queries using a suite of vision algorithms, focusing on selecting the most optimal algorithm through an adaptive query optimization approach.
Similarly, Chameleon~\cite{chameleon_jiang_2018} selects the most optimal algorithm through sliding window-based profiling.
While ExSample, FiGO, and Chameleon illustrate the necessity of adaptive query processing in VDBMSs, 
they propose custom query execution frameworks with limited generalizability and extensibility.
Additionally, these systems trade-off accuracy for faster queries, while \sys focuses on system-level optimizations that do not affect the accuracy of query results.
 
\subsection{AQP in Relational DBMS}\label{sec:back:aqp}
\begin{figure}[t]
\begin{subfigure}[t]{\linewidth}
  \centering
  \vskip 0pt
  \includegraphics[width=0.85\columnwidth]{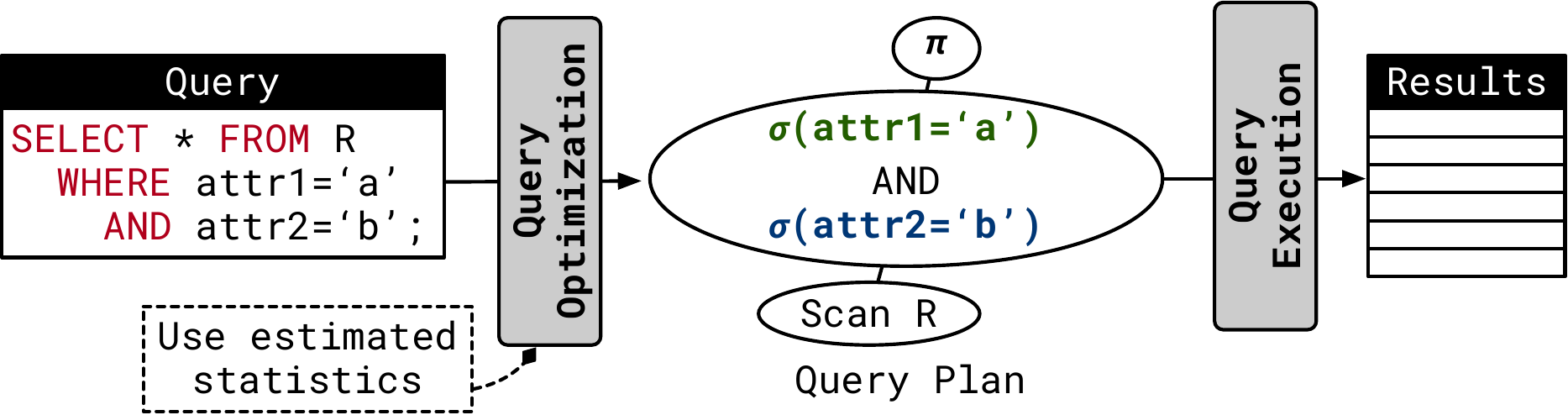}
  \caption{Static query processing pipeline.}
  \label{fig:back:aqp:base}
\end{subfigure}
\hfill
\begin{subfigure}[t]{\linewidth}
  \centering
  \includegraphics[width=0.85\columnwidth]{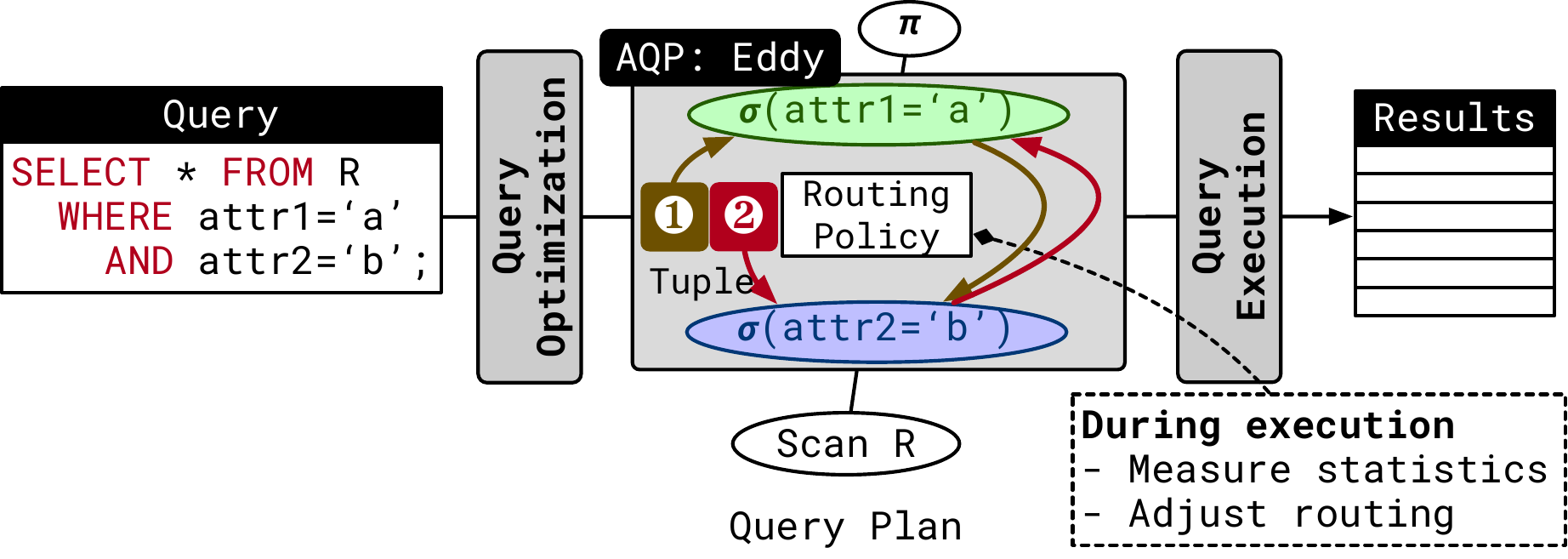}
  \caption{\aqp query processing pipeline.}
  \label{fig:back:aqp:eddy}
\end{subfigure}
\caption{\textbf{Query Execution Pipelines --}
In static query processing, the predicate ordering is determined based on statistics estimated during query optimization. 
In contrast, adaptive query processing dynamically governs the predicate ordering during query execution.} 
\end{figure}
A static query execution pipeline is commonly used by relational DBMSs, which consists of query parsing, query optimization, and query execution as shown in~\cref{fig:back:aqp:base}.
In this example, the query optimizer uses the estimated statistics to decide the optimal predicate ordering -- whether to evaluate the predicate on attribute one or attribute two first.
The estimated statistics can be inaccurate, often leading to sub-optimal query plans.
To tackle the shortcomings mentioned above, many prior efforts~\cite{aqpsurvey_deshpande_2007,midquery_kabra_1998,eddies_avnur_2000,contentaqp_babu_2005,proactive_babu_2005} have proposed an adaptive query processing scheme, that leverages statistics profiled at runtime to adjust the query plan.

\PP{\eddy}
Among these efforts, \eddy~\cite{eddies_avnur_2000} is a pioneer in proposing a systematic AQP framework that continuously reorders the application of pipeline operators in a query plan on a tuple-by-tuple basis.
The query optimizer constructs an \eddy operator during optimization.
For our example shown in~\cref{fig:back:aqp:eddy}, the \eddy operator contains two inner operators (a.k.a selection predicates), $\sigma\text{(R.attr1)}=\text{`a'}$ and $\sigma\text{(R.attr2)}=\text{`b'}$.
The routing policy inside the \eddy will determine which operator to evaluate first during execution.
\eddy will also monitor execution statistics like execution cost and selectivity to adjust the routing table for optimal routing.
In the original paper~\cite{eddies_avnur_2000}, routing policy based on both execution cost and predicate selectivity demonstrates good performance.
Intuitively, the predicate that runs fast and filters many tuples for later operators is prioritized. 
\eddy maintains an input queue for each inner operator, so the execution cost is inferred from the average queue length.
It also uses a lottery system~\cite{lottery_waldspurger_1994} to infer the selectivity of each inner operator.

\PP{Content-based Routing}
Babu~\etal~\cite{contentaqp_babu_2005} discover that the routing based on average statistics, as suggested in the original \eddy paper, can be significantly improved by doing data or content-based routing.
Their mechanism is built on top of the \eddy adaptive query processing framework.
The key idea is that a tuple could have some attributes that strongly correlate with the predicate selectivity.
By examining the value of those attributes, the \eddy operator can determine the optimal predicate ordering.
Nevertheless, the routing overhead is very high since the routing decision is made at tuple granularity.
\begin{figure*}
	\centering
	\includegraphics[width=0.85\textwidth]{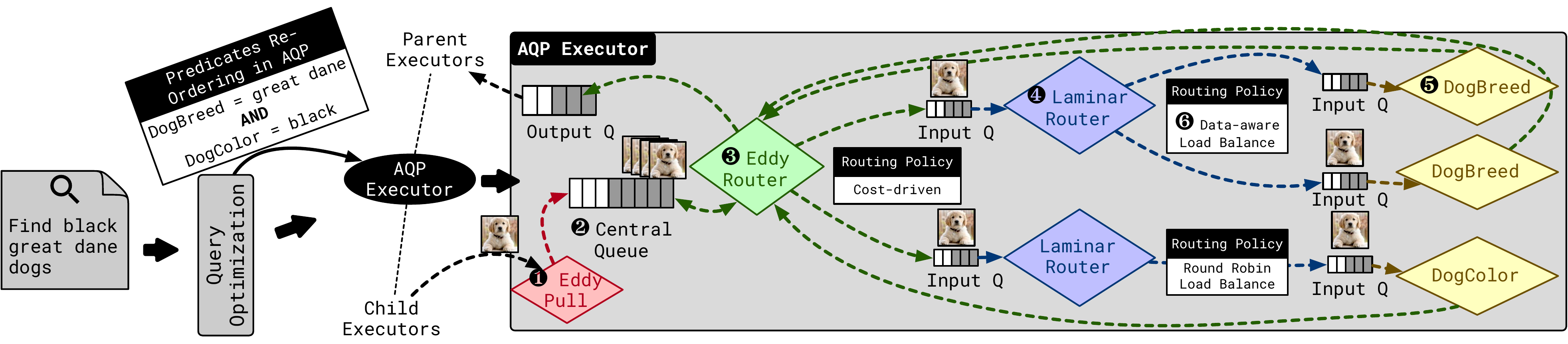}
	\caption{\textbf{Detailed \aqp execution plan and its internal} -- Left shows the execution tree with \aqp executor. \emph{Diamonds} represent physical processes apart from the main process. All physical queues serve as a medium for communicating between data producers and consumers. \emph{Rectangles} represent routing policies attached to according to processes.
	}
	\label{fig:impl:eddy_internal}
\end{figure*}
\section{AQP in \sys}\label{sec:impl}
We go over the design of \sys in this section.
\sys augments the \eva~\cite{kakkar_eva_2023}, an ML-centric database system, to support \aqp.
By enabling AQP, the goal is to achieve an optimal order of predicate execution and good scalability during query execution.
We first introduce how we integrate \aqp in \eva using the illustrative query shown in~\cref{lst:intro:example}.
We next describe the internals of the \aqp executor. 

\subsection{System Design}
\label{sec:impl:integrate}
\eva is a database system that optimizes and executes queries with \udfs powered by ML algorithms.
Like many other traditional database systems, \eva performs query parsing, optimization, planning, and execution. 
It supports static query optimization using a classical Cascades-style optimizer~\cite{graefe_cascades_1995}. 
With \sys, the query optimizer automatically relies on \aqp for handling ML-centric predicates for efficient query execution.

In the illustrative example (\cref{fig:impl:eddy_internal}), the query goes through regular parsing, optimization, and planning stages. 
Because the cost and selectivity of \udf are usually unknown during static optimization, \sys only performs rule-based optimizations like predicate pushdown, trivial predicate reordering\footnote{Predicates that do not involve \udfs}, and caching and reuse of results from resource-intensive \udfs \cite{xu_eva_2022}.
On the other hand, on the example query, the optimizer needs to handle \udf-based predicates for checking the color and breed of dogs.
Instead of doing sub-optimal query optimization, the query optimizer automatically constructs an \aqp plan for those two predicates, governed by the optimizer's rules. 
This \aqp plan is associated with the \aqp executor, which is responsible for generating an optimized predicate execution plan, dynamically reordering predicates, and scaling query workers during the query execution time (\cref{sec:impl:aqp}).

\subsection{\aqp Executor}\label{sec:impl:aqp}
Once the query plan transitions to an execution tree, the query executor executes in a top-down fashion.
The parent executor of \aqp pulls a batch of data from its output queue, while the \aqp executor pulls data from its child and performs all heavy computation asynchronously. 
At a high level, there are two key components in the \aqp executor: \eddy and \laminar routing. 
\eddy routing dictates the predicates' evaluation order. 
On the other hand, \laminar routing dictates the degree of parallelization. 
Since these predicates contain \udfs and are hence expensive, it spawns multiple instances of the \udfs to maximize hardware utilization with good load balancing. 
\laminarrouter controls how many such workers spawn and how to distribute the load between them. 
Since one batch is only required to be evaluated by one worker, we call it \laminar routing (inspired by the \laminar flow in fluid dynamics).

\PP{\aqp Internals}
We provide brief descriptions of the internal of \aqp executor following the example query (\cref{fig:impl:eddy_internal}).
\squishlist
\item[\ding{182}] \eddypull is a worker that pulls data batches from the child executor and inserts them into the \centralqueue. In this example, \eddypull gets bounding boxes that contain dogs from the downstream executor.
\item[\ding{183}] \centralqueue serves as a buffer for incoming batches from the child executor and batches after predicate computation.
\item[\ding{184}] \eddyrouter gets data from the \centralqueue and is responsible for orchestrating the data flow within the \aqp executor. 
In this example, it would prioritize scheduling data to the \lstinline[style=SQLStyle]{DogColor} predicate first due to its lower cost.
It also redirects complete data batches to the output queue or remaining unfinished predicates.
\item[\ding{185}] \laminarrouter gets data from \eddyrouter.
It is responsible for monitoring the hardware usage (\eg GPU utilization), determining the number of workers to spawn, and performing load balancing between workers.
Each predicate is associated with one \laminarrouter, so the executor constructs two \laminarrouter for \lstinline[style=SQLStyle]{DogBreed} and \lstinline[style=SQLStyle]{DogColor} in this example.
Additionally, because \lstinline[style=SQLStyle]{DogBreed} is more computationally intensive, the \laminarrouter spawns more workers after obtaining hardware usage of the \udf.
It uses a more advanced data-aware load-balancing mechanism for those two workers.
\item[\ding{186}] Spawned workers by the \laminar evaluate the predicate. 
The \laminarrouter spawns two and one workers for \lstinline[style=SQLStyle]{DogBreed} and \lstinline[style=SQLStyle]{DogColor} predicates, respectively.
After evaluating a data batch, it is inserted back to the \centralqueue.
\item[\ding{187}] Another important component of \aqp internal is each router's routing policy.
These policies govern the internal routing of batches within the \aqp executor. 
In the case of \eddy, it relies on the property of different predicates to determine its execution order.
In this example, it routes batch based on the cost of predicates and prioritizes the \lstinline[style=SQLStyle]{DogColor} predicate.
Conversely, the \laminarrouter{'s} objective is mainly to balance the load between workers.
For example, it can use round robin for simple load balancing, while it can also choose more advanced data-aware load balancing when the load is highly dependent on the characteristics of data batches. 
\squishend
\PP{Data Flow and Routing}
As mentioned earlier, the \eddypull retrieves batches from the child executor and inserts them into the \centralqueue.
In this example, data batches that the \eddypull gets from its child executor are bounding boxes that contain dogs.
The \eddyrouter orchestrates the data flow of the internal \aqp executor.
It fetches data from the \centralqueue and routes batches according to its routing policy to input queues of the \laminarrouter.
Once the data batch is in the input queue of the \laminar process, the \laminarrouter dictates further routing.
Based on its routing policy, the \laminarrouter selects workers and inserts data batches into their input queues. 
After the predicate worker evaluates the batch, they send data batches back to the \centralqueue.
Data batches that do not satisfy the predicate condition are dropped immediately.
In this example, data batches that are sent back to \centralqueue are bounding boxes that contain black great dane dogs.
Finally, the \eddyrouter directs data batches from the \centralqueue to the output queue once they complete all predicates.
To determine which predicates are already evaluated on a batch, the \eddyrouter maintains additional metadata about each batch. 
Further details about this metadata are provided in \cref{sec:impl:design}. 
The parent executor of the \aqp executor will pull data batches from the output queue of \aqp executor in a blocking way.
In this example, the parent executor is simply a projection that will display information about bounding boxes that contain black-colored lost dogs.

\subsection{Design Decisions}\label{sec:impl:design}
We next go over some design decisions in the \aqp executor.

\PP{Metadata for Data Routing}
The \aqp executor requires associating metadata with each batch. 
Therefore, to uniquely identify each batch, it assigns a unique id to every routing batch inserted into the \centralqueue by the \eddypull. 
This approach proves more efficient than using computationally expensive data hashing, particularly considering that the batches may contain large multi-dimensional data.
Furthermore, the \eddyrouter maintains a hash table to track predicates visited by the routing batch, utilizing its unique id.
This generic metadata is needed regardless of the routing policy used in the \eddyrouter.
Before routing the batch, the \eddyrouter checks the hash table to decide whether to skip or run the predicate, depending on its visitation status, thereby preventing redundant computation.
Once a data batch has visited all predicates, it is deemed a completed batch and subsequently routed to the output queue.

Moreover, the \eddyrouter can also maintain additional metadata for different routing policies.
For example, the cost-driven routing policy (shown in~\cref{fig:impl:eddy_internal}) necessitates monitoring input queue length and execution time for each predicate as part of its cost.
As required by the routing policy, the \eddyrouter can monitor and track these statistics as additional metadata, subsequently utilizing them to update routing decisions.

\PP{Eager Materialization}
During predicate evaluation, we employ an eager materialization approach, where the routing batch promptly discards tuples that it is certain do not satisfy the predicate condition. 
This strategy simplifies the predicate short-circuiting logic, removing the necessity to track information at the granularity of each row in the batch.

\PP{Deadlock Prevention}
Because the routing of the completed batch also goes through the \centralqueue, it becomes a point of resource contention.
If the \centralqueue is filled with batches that need to be sent to the \laminarrouter, the completed batch from the predicate workers cannot be reinserted into the \centralqueue, resulting in a deadlock for the entire \eddy executor.

To prevent deadlock, the \eddypull adopts a conservative approach when inserting batches into the \centralqueue.
Insertion only occurs if the \centralqueue is less than $\lambda$ percentage full. 
For our experiments, we set $\lambda = 0.3$.
Additionally, we configure the input queues for \laminarrouter and predicate workers to have very short lengths (\eg 2) to prevent the accumulation of too many backlog batches at those queues.

\PP{Batch Evaluation}
In the \sys, data is organized into a unit known as a routing batch, which helps amortize overhead in the pipeline (e.g., queue). 
In the current configuration, we have set the routing batch to include $10$ rows of data, although users have the flexibility to configure this number as needed. 
\section{USE CASES OF \eddy}\label{sec:eddy-uc}
In this section, we demonstrate multiple scenarios in video analytics that gain significant performance gains through usage of the adaptive query processing techniques presented in \sys.

\subsection{UC1: Cost-Driven Routing}\label{sec:eddy-uc:uc1}
For the first case, we aim to demonstrate that \sys can obtain accurate execution statistics for different {\udf}s and construct an optimal query plan based on that.

We consider an example in which a dog owner wants to identify their lost black great dane in a surveillance video taken at a park.
The owner formulates a query (\cref{lst:eddy-uc:uc1}) to retrieve frames of the surveillance video that might contain their lost dog by matching the breed and the color of observed dogs.

\begin{lstlisting}[style=SQLStyle, label=lst:eddy-uc:uc1, caption={Query to retrieve frames containing black great dane dogs.}]
SELECT id, bbox FROM video 
 CROSS APPLY UNNEST(ObjectDetector(frame)) AS Object(label, bbox, score) 
 WHERE Object.label='dog' 
 AND DogBreedClassifier(Crop(frame, bbox)) = 'great dane' 
 AND DogColorClassifier(Crop(frame, bbox)) = 'black';
\end{lstlisting}
The \lstinline[style=SQLStyle]{ObjectDetector} function returns the identified object category, its bounding box, and its classification score per frame. 
These results are flattened by \lstinline[style=SQLStyle]{UNNEST}, transforming the list of objects extracted from a frame into a set of object rows.
Subsequently, \lstinline[style=SQLStyle]{CROSS APPLY} associates each detected object row to its original video frame.
Then, the region that contains the object is cropped from the video frame using the \lstinline[style=SQLStyle]{Crop} \udf.
For the identified object regions, dog breed and color are further evaluated using \lstinline[style=SQLStyle]{DogBreedClassifier} and \lstinline[style=SQLStyle]{DogColorClassifier} {\udf}s.
In our implementation, we use state-of-the-art YoloV5~\cite{jocher_yolov5_2020} as \lstinline[style=SQLStyle]{ObjectDetector}, ViT~\cite{dosovitskiy_image_2021} finetuned on dog breed classification as \lstinline[style=SQLStyle]{DogBreedClassifier}, and a simple heuristic-based color classification by checking color value in the HSV space.

\begin{figure}[t]
\begin{subfigure}[t]{0.43\linewidth}
  \centering
  \vskip 0pt
  \includegraphics[width=\columnwidth]{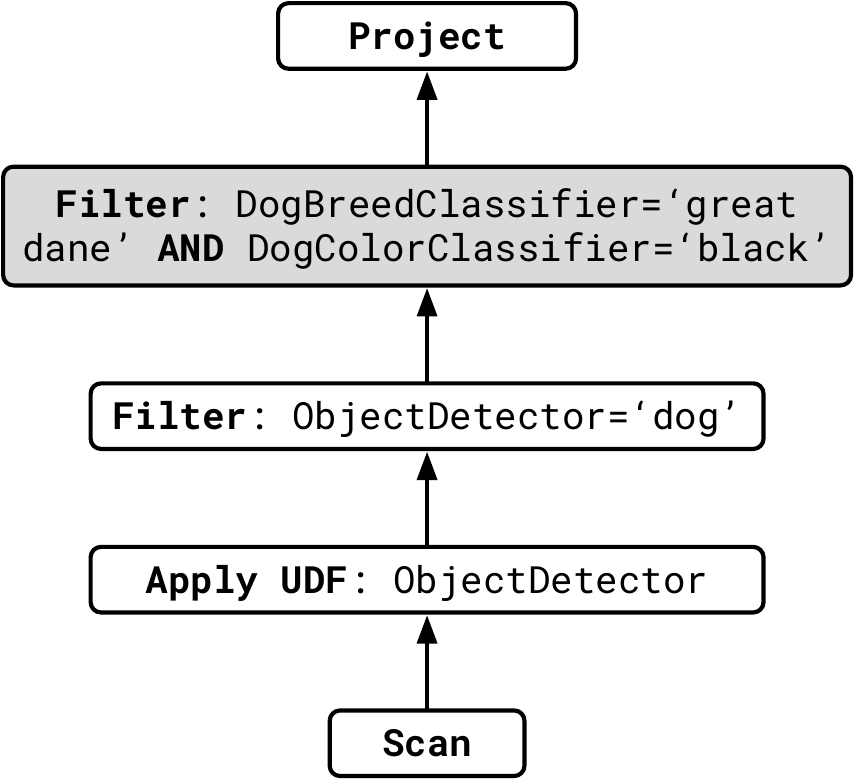}
  \caption{Query plan with no predicate reordering.}
  \label{fig:eddy-uc:uc1:base}
\end{subfigure}
\hfill
\begin{subfigure}[t]{0.52\linewidth}
  \centering
  \vskip 0pt
  \includegraphics[width=\columnwidth]{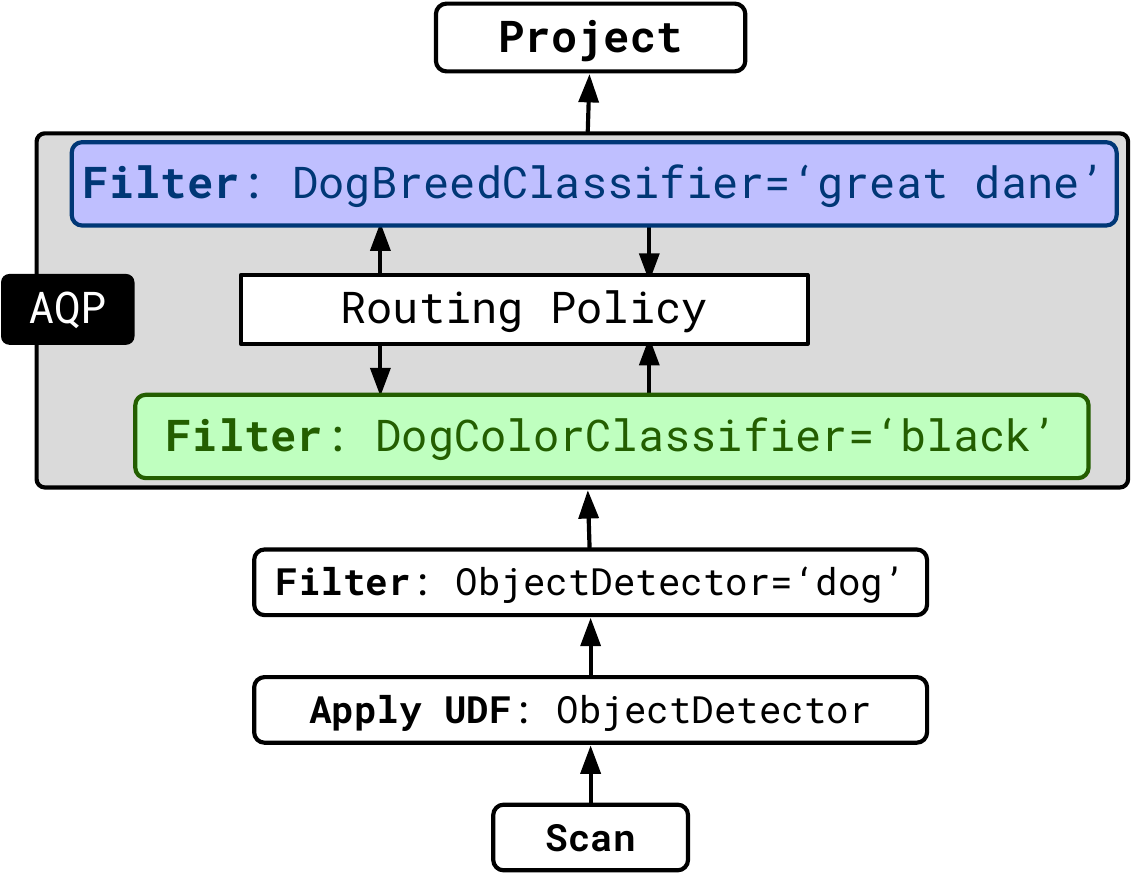}
  \caption{\aqp query plan that adaptively reorders predicate during query execution.}
  \label{fig:eddy-uc:uc1:eddy}
\end{subfigure}
\caption{\textbf{Query plan for UC1 --} query plan with and w/o predicate reordering for UC1.}
\label{fig:eddy-uc:uc1}
\end{figure}

\cref{fig:eddy-uc:uc1} illustrates the query plan without predicate reordering alongside the corresponding \aqp plan for the example query (\cref{lst:eddy-uc:uc1}).
The query plan first scans the video.
It then invokes the apply operator to run \lstinline[style=SQLStyle]{ObjectDetector}  on every frame.
The apply operator internally handles \lstinline[style=SQLStyle]{UNNEST} operations. 
Subsequently, the filter operator employs a simple predicate to extract all instances of dogs in the video.
The detected dog objects then pass through the filter operator, which executes the predicates by evaluating the \lstinline[style=SQLStyle]{DogBreedClassifier} and \lstinline[style=SQLStyle]{DogColorClassifier} \udfs. 
In the query plan without predicate reordering (\cref{fig:eddy-uc:uc1:base}), the \udfs inside the filter operator are executed in accordance with the conjunction order, proceeding from left to right.
In contrast, the \aqp query plan (\cref{fig:eddy-uc:uc1:eddy}) dynamically reorders the predicates during execution based on the routing policy.
Notably, the \lstinline[style=SQLStyle]{ObjectDetector} is excluded from the \aqp plan since it must be executed before the other two \udfs. Consequently, the room for further optimization is limited for \lstinline[style=SQLStyle]{ObjectDetector}.

\PP{Cost-Driven Routing Policy}
An optimal routing policy is crucial for the \aqp execution framework.
Previous studies exploring predicate reordering~\cite{eddies_avnur_2000,proactive_babu_2005,contentaqp_babu_2005} emphasize the significance of both the cost and selectivity of predicates for performance.
Ideally, a faster predicate capable of filtering a substantial amount of data is preferred to run first, leading to a significant reduction in the invocation of slower predicates. 
Consequently, a score function, $\frac{\text{cost}}{1 - \text{selectivity}}$, is commonly used to rank each predicate~\cite{hellerstein_practical_1994}.
The predicate with the lowest score is prioritized for execution first, contributing to optimal performance.

However, our findings (\cref{sec:eddy-uc:uc1-res}) indicate that relying on a scoring function for predicate reordering is not always optimal when dealing with concurrent workers.
For example, when one predicate is evaluated on the CPU and the other on the GPU, workers (\cref{fig:impl:eddy_internal}) associated with these predicates can run concurrently.
The same holds for scenarios where two predicates, each needing one CPU, run concurrently when the system has access to more than two CPUs. 
In concurrent settings, the empirical results (\cref{sec:eddy-uc:uc1-res}) show that the cost-driven routing policy delivers the same or outperforms the score-driven routing policy.

\begin{figure}[t]
  \centering
  \vskip 0pt
  \includegraphics[width=0.95\columnwidth]{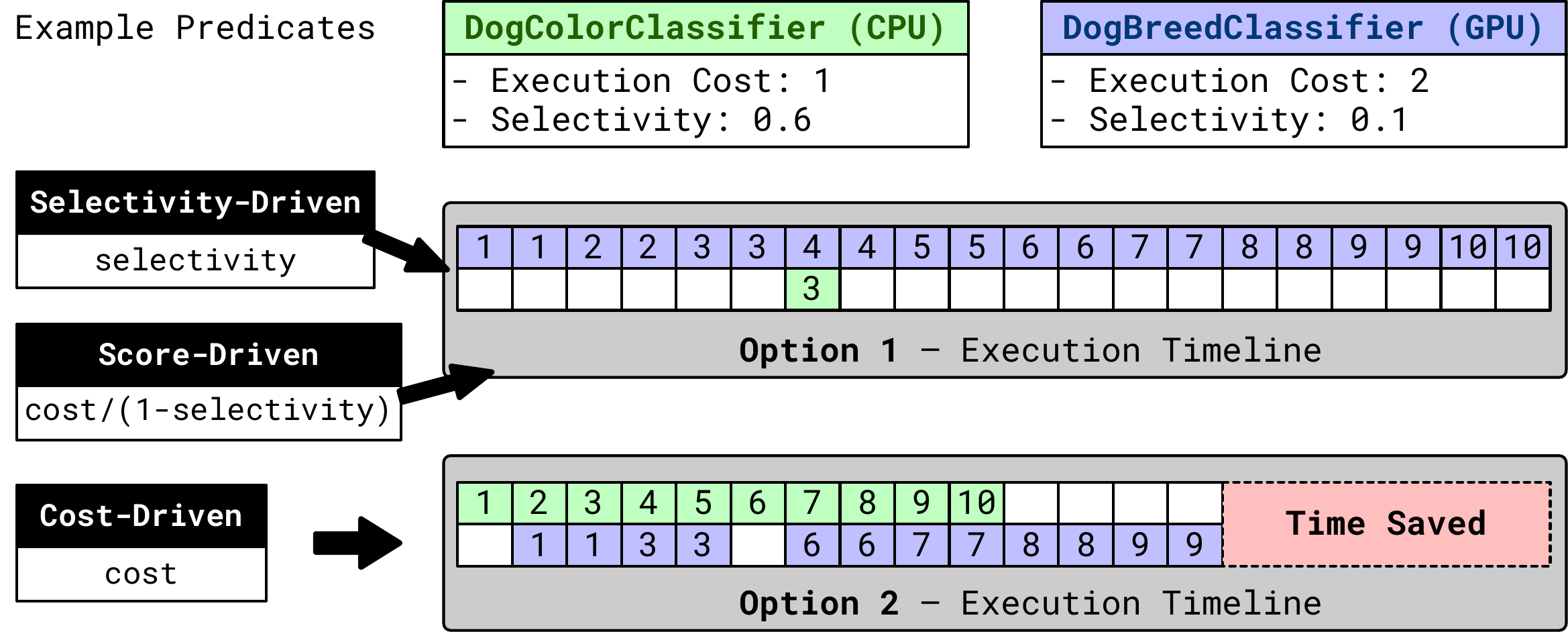}
  \caption{\textbf{Routing policy comparison --} execution timeline of selectivity-driven, score-driven, and cost-driven routing policy. One box represents a time unit.}
  \label{fig:eddy-uc:uc1:cost_route}
\end{figure}

\PP{Example comparison}
Our example query (\cref{lst:eddy-uc:uc1}) has two predicates: \lstinline[style=SQLStyle]{DogBreedClassifier} and \lstinline[style=SQLStyle]{DogColorClassifier}. 
The former, requiring a computationally expensive deep learning model, runs on the GPU, while the latter efficiently operates on the CPU. 
Consequently, these two predicates can run concurrently during query execution.
As mentioned earlier, in a concurrent setting, we notice a shift where the selectivity of a predicate no longer impacts query performance. 
Instead, only the execution cost of a predicate matters for optimal execution efficiency. 
We illustrate this shift in our reasoning using a simple example (\cref{fig:eddy-uc:uc1:cost_route}).

For both predicates in the example, we assign relative execution costs: \lstinline[style=SQLStyle]{DogColorClassifier} with a cost of $1$, and \lstinline[style=SQLStyle]{DogBreedClassifier} with a cost of $2$.
Despite running on the GPU, \lstinline[style=SQLStyle]{DogBreedClassifier} has a higher execution cost. 
Conversely, \lstinline[style=SQLStyle]{DogColorClassifier} runs on the CPU but has a lower execution cost.
We set the selectivity of \lstinline[style=SQLStyle]{DogColorClassifier} to $0.6$ and the selectivity of \lstinline[style=SQLStyle]{DogBreedClassifier} to $0.1$.
To examine the different routing policies, consider a total of $10$ units of data for evaluation.

For the selectivity-driven routing, \lstinline[style=SQLStyle]{DogBreedClassifier} is preferred over \lstinline[style=SQLStyle]{DogColorClassifier} as $0.1 < 0.6$. 
Similarly, for score-driven routing, \lstinline[style=SQLStyle]{DogBreedClassifier} is preferred over \lstinline[style=SQLStyle]{DogColorClassifier} as $\frac{2}{1 - 0.1} < \frac{1}{1 - 0.6}$.
In this case (first option), the execution spends $20$ time units on \lstinline[style=SQLStyle]{DogBreedClassifier} predicate, taking $2$ time units to process each data ($1$ box represents a time unit in \cref{fig:eddy-uc:uc1:cost_route}). 
Since only one out of ten data ($0.1$ selectivity) is passed to the \lstinline[style=SQLStyle]{DogColorClassifier} predicate for further evaluation, and it executes concurrently on CPU, the overall cost is $20$ time units.
In the cost-driven routing policy (second option), the execution spends $10$ units of time on \lstinline[style=SQLStyle]{DogColorClassifier}, and then six out of ten data points are passed to \lstinline[style=SQLStyle]{DogBreedClassifier} for further evaluation, resulting in a total time cost of $14$ units.
Despite the higher selectivity of \lstinline[style=SQLStyle]{DogColorClassifier}, leading to more data for further evaluation by the other predicate, the overall execution time is shorter due to better computation overlap between the two predicates.
The intuition is that since \lstinline[style=SQLStyle]{DogBreedClassifier} serves as the bottleneck in the pipeline due to its high execution cost, executing the faster predicate first helps alleviate the bottleneck, ultimately improving query performance.

Motivated by the above example, we choose a unique cost-driven routing policy in \sys when two predicates can run concurrently.
This approach sets \sys apart from other existing \aqp systems. 
For other situations where predicates must run on the same hardware resource and require synchronization between processes, \sys falls back to using the classic score-based approach.

\PP{Warmup Phase}
For predicates that run concurrently, it is important to route a batch of data to predicates based on their execution cost.
One issue faced by \sys is that the execution cost will not be available only after running a few initial batches of data.
To alleviate this issue, we introduce a warmup phase, in which batches are routed to all workers so that the exact execution cost of each predicate can be made available to the system.
After the warmup phase, batches are routed based on the chosen routing policy.
To ensure a low overhead of the warmup phase, we route just enough batches so that all predicates get executed.
To ensure all other remaining batches are routed in optimal execution order, we slightly delay the routing of other batches until the warmup phase completes (\ie the system finishes gathering accurate statistics for all predicates).
To achieve the delay routing during the system warmup and to avoid blocking the warmup batches in the central queue, we add a circular data flow, during which the delayed batches are pulled from the head of the central queue and inserted back to its tail.
In such a way, completed warmup batches eventually would reach the head of the central queue and mark the end of the warmup phase.

\subsection{UC1: Performance Results}\label{sec:eddy-uc:uc1-res}
In this section, we will show the performance benefit obtained using the \aqp technique with \sys.

\PP{Experimental Setup}
As mentioned, we implement \sys as part of \eva.
We use \eva $0.2+dev$ as our base framework.
We conduct experiments on a server with AMD EPYC $7452$ $32$-core processor with $256$ GB memory.
The server is also equipped with an NVIDIA A40 GPU, which has $48$ GB GPU memory.
The server runs a Ubuntu 22.04.3 LTS operating system, and the GPU library is compiled with NVIDIA CUDA 12.0.
We will maintain the same setup throughout the paper unless explicitly specified.
Throughout our paper, we employ \eva as our baseline, which, by default, utilizes a static optimization framework.
Therefore, the order of predicates is fixed for \eva during query execution.

\PP{Dataset}
We collected a video with various breeds of dogs with different colors from YouTube.

\PP{Predicate Implementation}
For \lstinline[style=SQLStyle]{DogColorClassifier}, we implement a simple heuristic, which classifies the object color based on the HSV range of common colors.
For example, the red color is defined within the range $(0, 50, 70)$ to $(9, 255, 255)$.
\lstinline[style=SQLStyle]{DogColorClassifier} can label colors as red, black, gray, yellow, green, blue, purple, pink, and white.
Colors that do not fall within the specified ranges are labeled as others.
For \lstinline[style=SQLStyle]{DogBreedClassifier}, we use a ViT-based transformer that is fine-tuned on dog breed classification tasks~\cite{yau_vit-dog_2023}.
It is capable of detecting a total of $120$ dog breeds.
We use the SotA YoloV5 model~\cite{jocher_yolov5_2020} as \lstinline[style=SQLStyle]{ObjectDetector}.

\begin{figure}[t]
  \centering
  \includegraphics[width=0.7\columnwidth]{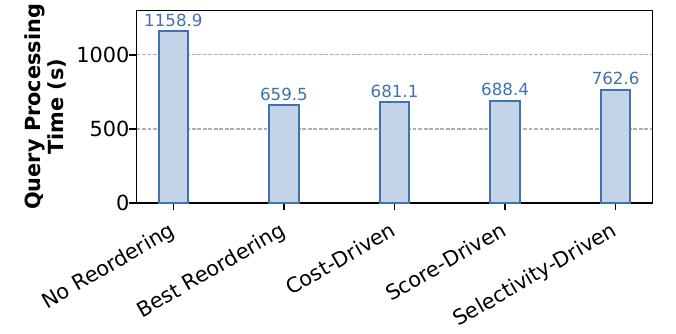}
  \caption{\textbf{Query processing time for UC1} -- comparison among five system options: no reordering, best reordering, \eddy cost-driven routing, \eddy score-driven routing, and \eddy selectivity-driven routing.}
  \label{fig:eddy-uc:uc1:perf}
\end{figure}

\PP{System Variants}
Throughout the paper, we extensively study five different system variants.
First, we introduce a \textsc{No Reordering} variant as our baseline, which refrains from predicate reordering during query optimization or execution, assuming that obtaining statistics for predicate reordering without the \aqp technique is challenging.
Second, we present the \textsc{Best Reordering} variant as the oracle for comparison.
For this variant, we profile statistics like predicate execution cost and selectivity ahead of time and perform predicate reordering manually before the query execution using the widely used predicate score-driven ranking function~\cite{hellerstein_practical_1994}: $\frac{\text{cost}}{1 - \text{selectivity}}$.
For both of the above variants, we disable the \aqp configuration setting. 
For \sys, we evaluate the performance of enabling the \aqp technique with three routing policies: \textsc{Cost-Driven}, \textsc{Score-Driven}, and \textsc{Selectivity-Driven}, prioritizing batch routing based solely on cost, $\frac{\text{cost}}{1 - \text{selectivity}}$, and solely selectivity, respectively.
For both score-driven and selectivity-driven routing methods, the \eddy router keeps track of the number of filtered rows for each predicate and calculates the selectivity based on that.

\PP{Benefit of Cost-Driven Routing}
We show the query processing time in~\cref{fig:eddy-uc:uc1:perf}.
We observe that \sys runs faster than no predicate reordering.
Cost-driven, score-driven, and selectivity-driven routing offer $1.70$\X, $1.68$\X, and $1.52$\X speedup, respectively.
Our profiling shows that the cost of evaluating \dogbreedudf is $35.11$ ms per tuple, and its selectivity is $0.254$.
While the cost of evaluating \lstinline[style=SQLStyle]{DogColorClassifier} is only $1.98$ ms per tuple, its selectivity is much higher ($0.633$).

During query processing, both cost-driven and score-driven routing policies choose to run \lstinline[style=SQLStyle]{DogColorClassifier} first, because it is faster compared to the other predicate.
On the other hand, the selectivity-driven routing policy schedules to run \lstinline[style=SQLStyle]{DogBreedClassifier} first because it is more selective.
As a result, selectivity-driven routing incurs a slightly higher query processing time ($762.6$ s) compared to the other two routing policies.
Cost-driven and score-driven have comparable performance.
Lastly, we observe that the query processing time for best reordering is $659.5$ s, which is comparable to the optimal options with \aqp enabled.
This verifies that both cost-driven and score-driven routing algorithms provide optimal predicate ordering during execution. 
The additional overhead of executing queries in \sys arises from startup processes (e.g., GPU context initialization in a new worker process).

\PP{Sensitive Analysis with Query Variants}
Next, we conduct a sensitivity analysis for different predicates, as demonstrated in~\cref{tb:eddy-uc:uc1:sensitive}.
We keep the same query as~\cref{lst:eddy-uc:uc1} but vary the predicate conditions. 
We consider two cases shown in~\cref{tb:eddy-uc:uc1:sensitive}.
Case $1$: high-cost predicate (\lstinline[style=SQLStyle]{DogBreedClassifier}) has significantly lower selectivity than the low-cost predicate.
Case $2$: high-cost predicate (\lstinline[style=SQLStyle]{DogBreedClassifier}) also has higher selectivity.

\begin{table}[t]
  \caption{\textbf{Charcteristics summary of different predicates} -- detailed statistics including selectivity and cost.}
  \small

  \resizebox{\columnwidth}{!}{
  \begin{threeparttable}

    \renewcommand{\arraystretch}{1.1}
    \centering
    \begin{tabular}{@{}c|ccc@{}}
    
    \toprule

    \textbf{Case}
    & \textbf{Predicate}
    & \textbf{Avg. Selectivity}
    & \textbf{Avg. Cost (ms)} \\

    \midrule \midrule
    
    \multirow{2}{*}{\textbf{1}}
    & \lstinline[style=SQLStyle]{DogBreedClassifier = 'Labrador retriever'}
    & 0.060
    & 29.516 \\

    & \lstinline[style=SQLStyle]{DogColorClassifier = 'Other'}
    & 0.374
    & 2.281 \\

    \midrule
    
    \multirow{2}{*}{\textbf{2}}
    & \lstinline[style=SQLStyle]{DogBreedClassifier = 'Great dane'}
    & 0.227
    & 28.315 \\

    & \lstinline[style=SQLStyle]{DogColorClassifier = 'Gray'}
    & 0.056
    & 1.974 \\
    
    \bottomrule
    
    \end{tabular}

  \end{threeparttable}
  }

  \label{tb:eddy-uc:uc1:sensitive}
\end{table}

\begin{figure}[t]
\begin{subfigure}[t]{0.49\columnwidth}
  \centering
  \includegraphics[width=\columnwidth]{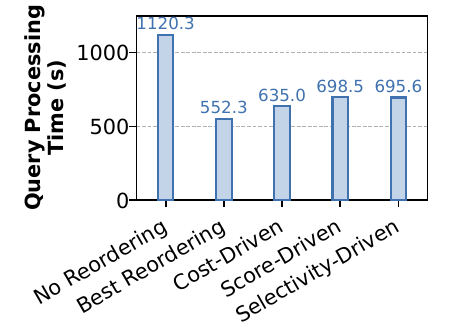}
  \caption{Low selectivity high-cost predicate.}
  \label{fig:eddy-uc:uc1:sen:case1}
\end{subfigure}
\hfill
\begin{subfigure}[t]{0.49\columnwidth}
  \centering
  \includegraphics[width=\columnwidth]{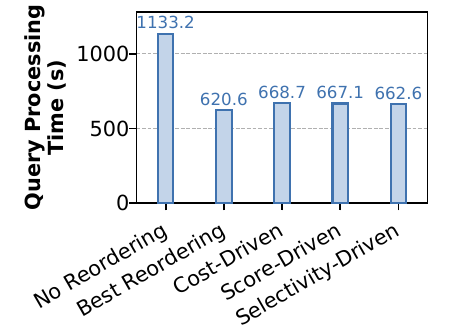}
  \caption{High selectivity high-cost predicate.}
  \label{fig:eddy-uc:uc1:sen:case2}
\end{subfigure}
\caption{\textbf{Routing policy sensitive study} -- query processing time comparison between different predicates with different selectivity and cost.}
\label{fig:eddy-uc:uc1:sen}
\end{figure}

\begin{figure}[t]
\begin{subfigure}[t]{\columnwidth}
  \centering
  \includegraphics[width=\columnwidth]{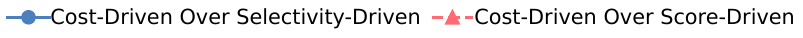}
\end{subfigure}
\hfill
\begin{subfigure}[t]{0.32\columnwidth}
  \centering
  \includegraphics[width=1.1\columnwidth]{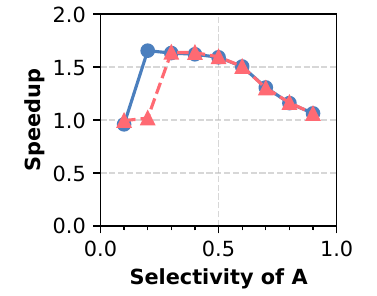}
  \caption{\textbf{\textit{B}: selectivity = $0.1$}}
  \label{fig:eddy-uc:uc1:syn-sen:case1}
\end{subfigure}
\hfill
\begin{subfigure}[t]{0.32\columnwidth}
  \centering
  \includegraphics[width=1.1\columnwidth]{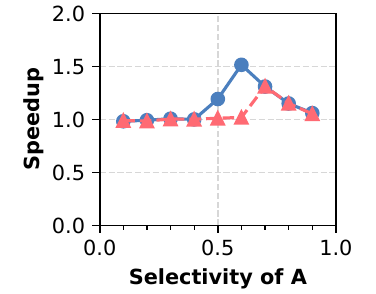}
  \caption{\textbf{\textit{B}: selectivity = $0.5$}}
  \label{fig:eddy-uc:uc1:syn-sen:case2}
\end{subfigure}
\hfill
\begin{subfigure}[t]{0.32\columnwidth}
  \centering
  \includegraphics[width=1.1\columnwidth]{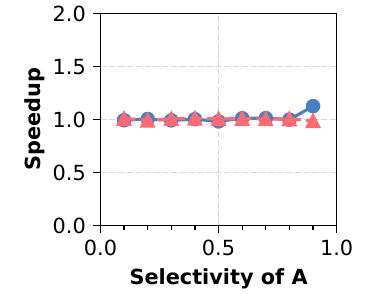}
  \caption{\textbf{\textit{B}: selectivity = $0.9$}}
  \label{fig:eddy-uc:uc1:syn-sen:case3}
\end{subfigure}
\caption{\textbf{Synthetic experiment} -- comparison to show the speedup of the cost-driven routing policy over others under different selectivities.}
\label{fig:eddy-uc:uc1:syn-sen}
\end{figure}

\cref{fig:eddy-uc:uc1:sen} presents the query processing time for both queries. 
The results demonstrate that all different routing options exhibit a notable speedup compared to no predicate reordering.
For the first case (\cref{fig:eddy-uc:uc1:sen:case1}), the query processing time for the cost-driven routing policy is slightly lower than both the selectivity-driven and score-driven routing policies.
We observe that even though the score-driven routing policy should have chosen the \lstinline[style=SQLStyle]{DogColorClassifier} predicate first based on the overall cost and selectivity, statistics fluctuation during query execution can cause it to route in a suboptimal order for some segments of the data.

In the second case (\cref{fig:eddy-uc:uc1:sen:case2}), all routing options of \sys achieve a similar query processing time.
For this case, \lstinline[style=SQLStyle]{DogColorClassifier} is chosen to process data first by all the policies.
Lastly, the cost-driven routing has comparable query processing time as the best reordering for both cases. 
The additional time of cost-driven routing is caused by the startup overhead and queuing delay.

\PP{Sensitive Analysis using Synthetic Queries}
For this experiment, we conduct a thorough investigation comparing the performance of cost-driven routing with both score-driven and selectivity-driven routing.
For this purpose, we define two predicates, denoted as \textit{A} and \textit{B}, with execution costs of $10$ ms and $20$ ms.
The selectivity of predicate \textit{B} is configured to be $0.1$, $0.5$, and $0.9$ as shown in~\cref{fig:eddy-uc:uc1:syn-sen:case1},~\cref{fig:eddy-uc:uc1:syn-sen:case2}, and~\cref{fig:eddy-uc:uc1:syn-sen:case3}, respectively.
We then vary the selectivity of \textit{A} from $0.1$ to $0.9$.
We demonstrate the query processing speedup of the cost-driven routing policy over both the score-driven and the selectivity-driven routing policies.
The results show that the cost-driven routing policy never provides a worse query processing time than the other routing policies.
Moreover, the cost-driven routing policy tends to perform better than the score-driven and selectivity-driven routing policy when the high-cost predicate has low selectivity.
Among all routing policies, the solely selectivity-driven routing policy exhibits the worst performance.
 
\subsection{UC2: Adaptive Routing}\label{sec:eddy-uc:uc2}
In the use case 1, we study the \sys{'s} cost-driven routing, which leverages the \eddy's approach to measure the cost of predicates and determine the appropriate order during the execution.
Since the cost of the predicates does not change during the execution, the optimal ordering of the predicates and, consequently, the query plan remains fixed.
However, our exploration of real-world exploratory analysis queries reveals that the cost of a particular predicate can significantly change during execution due to other optimizations or variations in data characteristics.
This motivates us to examine the benefits of using \sys for execution-time adaptive routing.
\begin{lstlisting}[style=SQLStyle, label=lst:eddy-uc:uc2, caption={Query to identify unsafe situation in warehouse.}]
-- Q1: Initial exploratory query
SELECT id, ObjectDetector(data).labels FROM video 
 WHERE id > 1000 AND id < 7000;

-- Q2: Initial exploratory query
SELECT id, HardHatDetector(data).labels FROM video 
 WHERE id > 8000 AND id < 14000;

-- Q3: Recurrent query
SELECT id FROM video
 WHERE ['person'] <@ ObjectDetector(data).labels
 AND ['no hardhat'] <@ HardHatDetector(data).labels;
\end{lstlisting}

\PP{Motivating Example}
~\cref{lst:eddy-uc:uc2} illustrates a motivating example of exploratory data analysis that involves multiple queries exploring video footage from a construction site.
For ease of demonstration, we consider simplified predicate conditions for the queries. 
In equivalent real-world scenarios, users may be interested in conducting deeper analyses of video segments based on factors such as time, weather conditions, working locations, etc.

In $Q1$, the user examines objects in the video between frame ids $1000$ and $7000$ using \lstinline[style=SQLStyle]{ObjectDetector}.
Next, in $Q2$, the user focuses on identifying hard hats using \lstinline[style=SQLStyle]{HardHatDetector} between frame ids $8000$ and $14000$.
Subsequently, in $Q3$, they run a query involving the previous \udfs to identify unsafe situations in the video where workers are not wearing hard hats when they should be.
Given that evaluating the same \udf on the same data across queries is a common occurrence in real-world exploratory use cases, EvaDB~\cite{xu_eva_2022} incorporates optimizations such as result caching and reusing to expedite these queries.
In this scenario, the outcomes of executing \lstinline[style=SQLStyle]{ObjectDetector} for the range \lstinline[style=SQLStyle]{id > 1000 AND id < 7000} and \lstinline[style=SQLStyle]{HardHatDetector} for the range \lstinline[style=SQLStyle]{id > 8000 and id < 14000} are cached after the execution of $Q1$ and $Q2$.
Consequently, when $Q3$ is executed, it efficiently reuses the cached results.

Note, for range \lstinline[style=SQLStyle]{id > 1000 AND id < 7000}, the cost of \lstinline[style=SQLStyle]{ObjectDetector} will be significantly lower than \lstinline[style=SQLStyle]{HardHatDetector} because cached results will be reused which eliminates the need for running the \lstinline[style=SQLStyle]{ObjectDetector} again.
Therefore, the optimal query plan should prioritize \lstinline[style=SQLStyle]{ObjectDetector} predicate for this range.
Similarly, the optimal plan should instead prioritize \lstinline[style=SQLStyle]{HardHatDetector} for the range \lstinline[style=SQLStyle]{id > 8000 and id < 14000} because \lstinline[style=SQLStyle]{HardHatDetector} will reuse cached results.
In conclusion, the use of partial caches introduces variability in the optimal predicate ordering during query execution.

\PP{Reuse-aware Routing}
To enhance our routing logic, we propose an \reuseaware routing algorithm built upon the cost-driven routing~\cref{sec:eddy-uc:uc1}.
In addition to the statistics collected for cost-driven routing, we also incorporate cache hit statistics.
The cache hit rate is utilized to determine the actual cost of a predicate when no cache is present.
During routing, the router algorithms first check the potential cache hit rate for a batch.
In our implementation, we utilize an on-disk key-value store for cache storage, allowing us to obtain an accurate cache hit rate per tuple in the batch with minimal overhead.
After getting the cache hit rate and the actual predicate cost, the routing algorithm estimates the potential execution cost for a routing batch using the following equation.
\[ \text{estimated cost} = \left(1 - \text{cache hit rate}\right) \cdot \text{cost of computing UDF}\]
We assume that the cache access overhead is negligible compared to the actual cost of the predicate. 
Lastly, the routing algorithm prioritizes scheduling data to the lowest-estimated-cost predicate to get the optimal performance.

\subsection{UC2: Performance Results}\label{sec:eddy-uc:uc2-res}
\PP{Experimental Setup}
To demonstrate the benefit of \reuseaware routing, we study the performance of the query shown in~\cref{sec:eddy-uc:uc2}.
For this example, we collect a video from YouTube where workers are operating in a warehouse.
In the video, some workers wear hard hats while others are operating without hard hats (\ie an unsafe situation).
For the \lstinline[style=SQLStyle]{ObjectDetector} predicate, we use the StoA YOLOv5 model.
For the \lstinline[style=SQLStyle]{HardHatDetector} predicate, we use the YOLOv8s model that is fine-tuned for hard hat detection.  
Before measuring the query processing time of the recurrent query $Q3$, we execute initial exploratory queries and cache the results.

\begin{figure}[t]
\centering
\includegraphics[width=0.55\columnwidth]{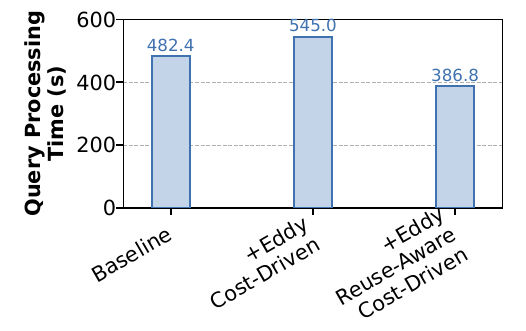}
\caption{\textbf{Query processing time for UC2} -- comparison among baseline, +\sys cost-driven, and +\sys \reuseaware cost-driven.}
\label{fig:eddy-uc:uc2:perf}
\end{figure}

\PP{Results}
In~\cref{fig:eddy-uc:uc2:perf}, we present the query processing times for the baseline, +\sys cost-driven, and +\sys \reuseaware cost-driven settings.
First, +\sys \reuseaware cost-driven routing achieves a speedup of $1.25 \times$ over the baseline, with a processing time of $386.81$ s compared to $482.41$ s. 
On the other hand, +\sys cost-driven only routing has a longer query processing time ($545.03$ s) than the baseline. 
Thus, +\sys \reuseaware cost-driven routing can provide $1.41$\X speedup compared to blindly using +\sys cost-driven only routing.
 
\begin{figure}[t]
\begin{subfigure}[t]{\columnwidth}
  \centering
  \includegraphics[width=0.5\columnwidth]{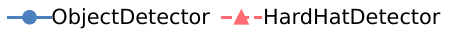}
\end{subfigure}
\hfill
\begin{subfigure}[t]{0.49\columnwidth}
  \centering
  \includegraphics[width=\columnwidth]{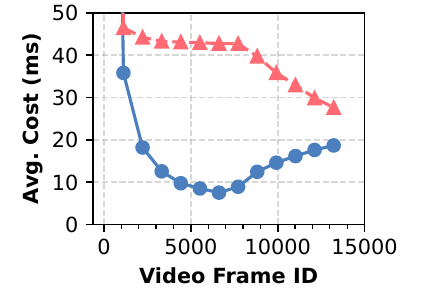}
  \caption{Cost-driven.}
  \label{fig:eddy-uc:uc2:plan:eddy}
\end{subfigure}
\hfill
\begin{subfigure}[t]{0.49\columnwidth}
  \centering
  \includegraphics[width=\columnwidth]{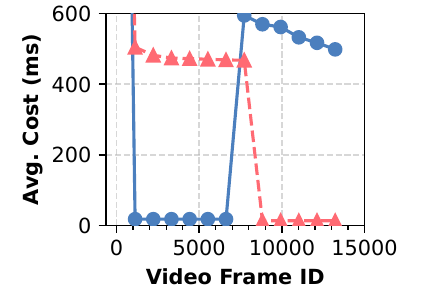}
  \caption{\reuseaware cost-driven.}
  \label{fig:eddy-uc:uc2:plan:eddy-opt}
\end{subfigure}
\caption{\textbf{Predicate average cost} -- the average cost of different predicates over video frame ID with different routing policies.}
\label{fig:eddy-uc:uc2:plan}
\end{figure}

\PP{Analyzing Estimated Predicate Cost}
To understand the results shown in results in~\cref{fig:eddy-uc:uc2:perf}, we examine the estimated costs of predicates across the two routing policies over the video (\cref{fig:eddy-uc:uc2:plan}).
Note that during routing, predicates with lower estimated costs are prioritized.
As shown in ~\cref{fig:eddy-uc:uc2:plan:eddy}, the estimated cost of \lstinline[style=SQLStyle]{ObjectDetector} dramatically decreases starting from frame \lstinline[style=SQLStyle]{id > 1000} to \lstinline[style=SQLStyle]{id < 7000}, because most of the results are cached.
Consequently, for this video range, data is directed to the \lstinline[style=SQLStyle]{ObjectDetector} predicate, representing the optimal plan.
However, because the cost-driven routing is not \reuseaware, the estimated cost cannot promptly adjust for the later part of the video (\ie \lstinline[style=SQLStyle]{id > 8000}).
As a result, the sub-optimal plan is used for the later part of the video, where data continues to be routed to the \lstinline[style=SQLStyle]{ObjectDetector} predicate first.

In contrast, \sys \reuseaware cost-driven routing is able to promptly adjust the estimated cost of the predicate for different segments of the video, as shown in~\cref{fig:eddy-uc:uc2:plan:eddy-opt}.
Specifically, the estimated cost of \lstinline[style=SQLStyle]{ObjectDetector} is adjusted to a very low value for the range of \lstinline[style=SQLStyle]{id > 1000 AND id < 7000} due to cache.
Likewise, the estimated cost of \lstinline[style=SQLStyle]{HardHatDetector} is also adjusted for the later part of the video.
Consequently, when \sys \reuseaware cost-driven routing is enabled, data is consistently routed to the lower-cost predicate, resulting in a better query execution plan compared to \sys cost-driven routing.

Last, we also observe that the baseline is slightly faster than \sys cost-driven in~\cref{fig:eddy-uc:uc2:perf}.
This is because the baseline setting always goes with the naive plan that has a fixed predicate order.
However, \sys cost-driven has a warmup phase when it starts, during which it routes data to all predicates regardless of their actual cost to gather some initial statistics about both predicates, this causes some data to be routed in a sub-optimal order compared to the baseline, so \sys cost-driven has slightly higher query processing time in this case.
\section{USE CASES OF LAMINAR}\label{sec:laminar-uc}
In the previous section, we explored two use cases that benefit from the \eddy operator, which determines the optimal predicate execution order based on their runtime statistics.
Beyond the order of predicate execution, another pivotal factor influencing query performance is the utilization of underlying hardware resources.
This is particularly crucial for operators and functions that heavily depend on GPUs (\eg \objdetudf).
For example, as shown in \cref{sec:laminar-uc:uc3-res}, underutilization of GPUs can degrade performance by a factor of $4.24 \times$.
To address this, we introduced the \laminar operator following the \eddy operator, illustrated in~\cref{fig:impl:eddy_internal}.
The \laminar operator enhances query performance by (1) ensuring optimal hardware utilization, particularly for GPU resources; (2) facilitating robust scalability as the system increases the number of resources; and (3) achieving effective load balancing among multiple backend workers during query execution.

\subsection{UC3: Hardware Utilization and Scalability}\label{sec:laminar-uc:uc3}
We showcase the features of the \laminar operator using~\cref{lst:laminar-uc:uc3} without the initial exploratory queries.
\begin{lstlisting}[style=SQLStyle, label=lst:laminar-uc:uc3, caption={Query to identify unsafe situation in warehouse without caches.}]
SELECT id FROM video
 WHERE ['person'] <@ ObjectDetector(data).labels
 AND ['no hardhat'] <@ HardHatDetector(data).labels;
\end{lstlisting}

\PP{Background}\label{sec:laminar-uc:uc3:hw}
The query's performance is significantly influenced by the utilization of the underlying hardware. 
In queries like~\cref{lst:laminar-uc:uc3}, where the GPU-intensive parts (\eg \lstinline[style=SQLStyle]{ObjectDetector}) are the bottleneck, the efficiency of GPU usage determines the final query execution performance.
Prior works in real-time serving for deep neural networks~\cite{xiao_gandiva_2018,dhakal_gslice_2020} emphasize the importance of GPU utilization for achieving high throughput in real-time model serving.
While most works~\cite{crankshaw_clipper_2017,romero_infaas_2021,shen_nexus_2019,wu_serving_2022,hu_scrooge_2021} aim to improve throughput without compromising the Service Level Objective (SLO) latency, they also introduce techniques to improve GPU utilization for maximal query throughput.
One of the most effective and straightforward of these techniques is adaptive run-time batching.
This method determines the ideal batch size during runtime, groups data (\eg tensors) into a batch, and performs DNN inference with the determined batch size.

\PP{Challenges}
However, the adaptive run-time batching doesn't entirely apply to our use case for two reasons.
\squishitemize
\item Firstly, batching assumes uniform dimensions across all input tensors, which is not always feasible in many practical applications.
\udf composition (\eg \lstinline[style=SQLStyle]{DogBreedClassifier(Crop(frame, bbox))}) is one such example.
In this case, each video frame is initially cropped based on the bounding box results, and the cropped region is analyzed by the \lstinline[style=SQLStyle]{DogBreedClassifier} predicate.
The dimension of inputs to the \lstinline[style=SQLStyle]{DogBreedClassifier} can vary significantly, so it is impossible to use the data batching technique for better GPU utilization.
\item Additionally, \sys heavily relies on \udfs interface to allow database users to use any third-party ML algorithms through \sys.
However, many third-party libraries may support single-batch inference as its most common interface.
For example, the YOLOv8 API~\cite{yolov8_ultralytics_2023} is commonly used, which exposes an API interface where users pass a single image as a parameter. 
This can lead to serious GPU under-utilization issues, necessitating a solution to overcome the fixed batch based on the actual GPU usage.
\squishend

\begin{figure*}
  \centering
  \vskip 0pt
  \includegraphics[width=0.95\textwidth]{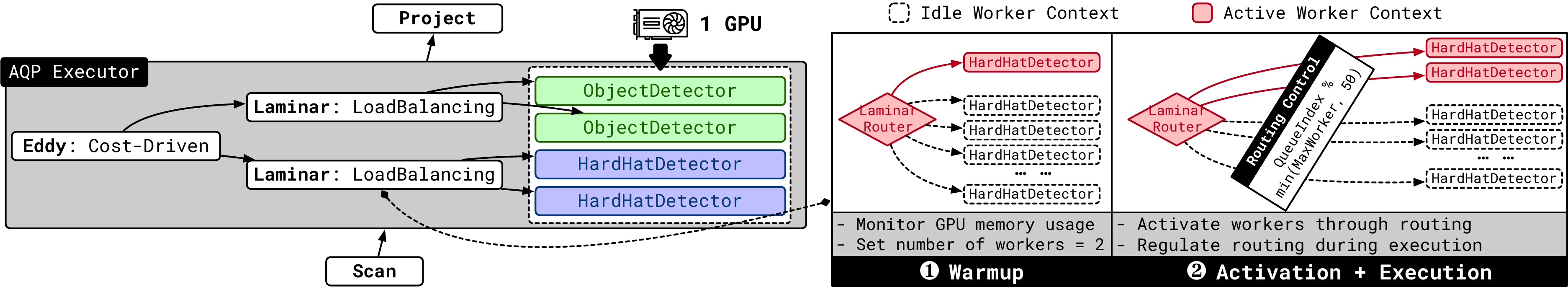}
  \caption{\textbf{Query plan for UC3} -- two-way parallelization enabled through \laminar with single GPU.}
  \label{fig:laminar-uc:uc3:plan}
\end{figure*}

\PP{Batch-agnostic Parallelization}
Inspired by the spatial-multiplexing approach from previous works~\cite{xiao_gandiva_2018,hu_scrooge_2021}, we implement a batch-agnostic parallelization approach for the \laminar operator.
The key difference is that each predicate worker only evaluates one batch of data, but there will be multiple predicate workers that perform evaluation simultaneously. 
The \laminar operator spawns multiple concurrent predicate workers when the GPU usage is low, allowing workers to overlap data movement, CPU computation, and GPU computation.

The query plan for~\cref{lst:laminar-uc:uc3} with two-way parallelization is depicted in~\cref{fig:laminar-uc:uc3:plan}.
In this example, the \laminar operator spawns two workers for each predicate, assuming the system has only access to one GPU.

\PP{GACU: Greedy-Allocation-Conservative-Use}
Even though the spatial-multiplexing approach has been well explored, we note that our key contribution is to support this feature in the context of \aqp framework.
The \laminar router is responsible for determining the parallelism and spawning workers during execution.
Ideally, maximizing the parallelism is desirable.
However, one critical constraint is the risk of out-of-memory errors when too many workers are allocated to the same GPU.
To mitigate this, the \laminar router can be designed to dynamically adjust the number of workers based on their memory usage. 
Nonetheless, dynamically expanding the worker count during execution poses challenges due to the need for the framework to acquire more processes, construct queues, and adjust the pipeline during query execution.

To tackle these challenges, we invent a straightforward yet effective approach called greedy-allocation-conservative-use (GACU), in which the \laminar router can allocate more workers through routing as needed during execution with no additional modification to the framework.
The core concept of GACU is to greedily allocate a considerable number of worker contexts when the query starts, but only activate a small subset of those workers during the execution based on runtime statistics (\eg GPU memory usage).
In our setting, we set a hardcoded value of $50$ worker contexts per GPU. 
It is worth noting that having $50$ workers per GPU exceeds the necessary amount for utilizing GPU resources during query execution. 
Typically, the number of workers that won't lead to GPU memory errors is much lower than this hardcoded number.

In our implementation, a worker context is a ray remote function and the \laminar router communicates data with the worker through ray queues.
Workers follow a lazy mechanism, avoiding proactive allocation of GPU resources until requests are present in their queues.
By doing so, only activated workers need to consume the GPU resources.
Additionally, the \laminar router no longer needs to adjust the query pipeline during the execution.
It can simply route data to workers to activate them as if those workers are being spawned to GPUs (\ie spawning through routing).

As demonstrated in~\cref{fig:laminar-uc:uc3:plan}, during warmup, the \sys conservatively only activates one worker. 
The \laminar router monitors the memory usage of this worker and determines the actual number of workers to activate based on this information.
After the warmup, the \laminar router updates the number of workers to be $\lfloor{\frac{\text{Total GPU memory}}{\text{Used GPU memory}}}\rfloor$.
It subsequently activates the remaining workers until the number of active workers reaches the determined threshold. 
Throughout query execution, the routing logic directs data exclusively to active workers, avoiding the activation of new ones.
Other not-activated worker contexts will stay idle until they are cleaned up at the end of the query execution.

\PP{Scaling Out}
When the system has access to multiple GPUs, all GPUs will be assigned workers to run the same predicates.
We follow a similar approach to determine the number of workers per GPU as discussed, whose upper-bound is also set to $50$.
The key difference is that the \laminar router now manages workers situated on different physical GPUs. 
To ensure a good query execution performance, we adopt a GPU-aware \laminar routing policy, in which we configure the routing logic to alternate between GPUs when routing a continuous data sequence.
Through our experiment, we find this approach can achieve a good load balancing between GPUs and consequently improve the overall GPU utilization.

\subsection{UC3: Performance Results}\label{sec:laminar-uc:uc3-res}
\begin{figure}[t]
\begin{subfigure}[t]{0.46\columnwidth}
	\vskip 0pt
	\centering
	\includegraphics[width=\columnwidth]{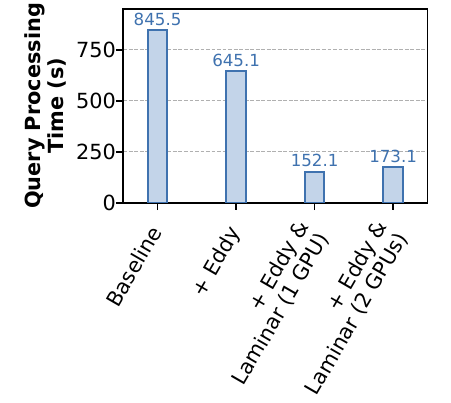}
    \caption{Short video}
    \label{fig:laminar-uc:uc3:perf:short}
\end{subfigure}
\begin{subfigure}[t]{0.53\columnwidth}
	\vskip 0pt
	\centering
	\includegraphics[width=\columnwidth]{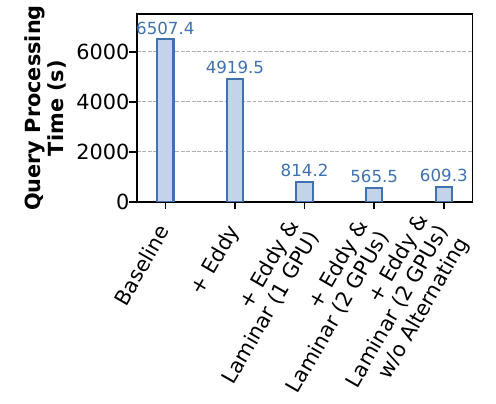}
    \caption{Long video}
    \label{fig:laminar-uc:uc3:perf:long}
\end{subfigure}
\caption{\textbf{Query processing time for \laminar UC3} -- comparison among different options: baseline, $+$ \eddy, $+$ \eddy \& \laminar (1 GPU), and $+$ \eddy \& \laminar (2 GPUs).}
\label{fig:laminar-uc:uc3:perf}
\end{figure}
We evaluate the benefit of using the \laminar router to improve the GPU utilization and the overall query execution time in this section.
To understand the benefit of each component better, we examine three options: 1) baseline; 2) $+$ \eddy cost-driven routing; and 3) $+$ \eddy cost-driven routing and \laminar routing on single GPU for our example query (\cref{lst:laminar-uc:uc3}) for both a short and a long video.
The long video is created by simply duplicating frames in the short video.
The longer video has a total of $112912$ frames, while the short video has $14114$ frames.
In addition to that, we also evaluate the $+$ \eddy cost-driven routing and \laminar routing on two GPUs to understand the scalability of this approach.

We show the query processing time in~\cref{fig:laminar-uc:uc3:perf}.
For the short video, the \eddy cost-driven routing takes a total of $645.11$ seconds to process the query, while the baseline takes $845.53$ seconds ($1.31$\X speedup).
Because the \eddy cost-driven routing does not optimize the GPU utilization, the $+$ \laminar routing option further reduces the query processing time to $152.10$ seconds, which is $5.56$\X and $4.24$\X speedup compared to the baseline and \eddy routing only.
While we expect running all optimizations (both \eddy and \laminar routings) will further reduce the query processing time, its total query processing is $173.13$ seconds, which is slightly higher than the single GPU option.
This is attributed to the short video, where worker startup overhead dominates the execution time.
For the long video, all optimizations together on a single GPU have a $7.99$\X speedup against the baseline, while the two GPU settings show $11.52$\X speedup.
Running the query on two GPUs now has a $1.44$\X speedup compared to a single GPU.
This demonstrates that our approach scales when the computation time is the dominant overhead.
We also added an additional option: $+$ \eddy \& \laminar (2 GPUs) w/o alternating in the experiment to show the benefit of doing GPU-aware routing (\ie alternating for load-balancing).
The results show that if GPU-aware routing is disabled, the total query processing increases from $565.51$ to $609.30$ seconds due to the load imbalance issue.

\begin{figure}[t]
\begin{subfigure}[t]{0.32\columnwidth}
  \centering
  \includegraphics[width=1.1\columnwidth]{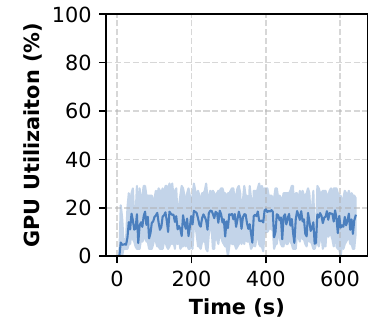}
  \caption{$+$ \eddy}
  \label{fig:laminar-uc:uc1:gpu-util:eddy}
\end{subfigure}
\hfill
\begin{subfigure}[t]{0.32\columnwidth}
  \centering
  \includegraphics[width=1.1\columnwidth]{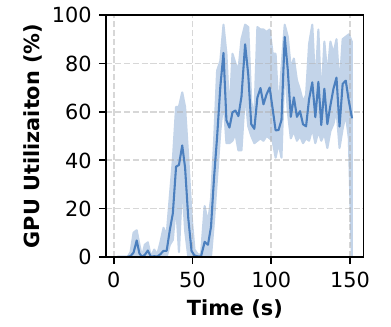}
  \caption{$+$ \eddy \& \laminar (1 GPU)}
  \label{fig:laminar-uc:uc1:gpu-util:laminar1}
\end{subfigure}
\hfill
\begin{subfigure}[t]{0.32\columnwidth}
  \centering
  \includegraphics[width=1.1\columnwidth]{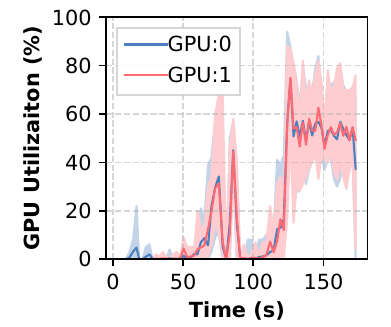}
  \caption{$+$ \eddy \& \laminar (2 GPUs)}
  \label{fig:laminar-uc:uc1:gpu-util:laminar2}
\end{subfigure}
\caption{\textbf{GPU utilization of different system options} -- the average, min, and max GPU utilization of different system options: $+$ \eddy, $+$ \eddy \& \laminar (1 GPU), and $+$ \eddy \& \laminar (2 GPUs).}
\label{fig:laminar-uc:uc1:gpu-util}
\end{figure}

\PP{GPU Utilization Study}
For the short video, we profile the utilization of GPUs for three system options shown in~\cref{fig:laminar-uc:uc1:gpu-util}.
These three figures illustrate the average GPU utilization (middle line), along with the minimum and maximum GPU utilization (depicted by the shaded region) over a windowed time.
We can see that if we only use the cost-driven routing, the average GPU utilization only stays around $20\%$ for the entire query execution period.
When we add the \laminar routing optimization, the GPU utilization is significantly improved for the single GPU (\cref{fig:laminar-uc:uc1:gpu-util:laminar1}).
Meanwhile, it also adds overhead to the system.
In \cref{fig:laminar-uc:uc1:gpu-util:laminar1}, we observe that the GPU goes through a period of low utilization at the beginning.
This occurs during the activation of workers, where GPU resources are allocated before the actual inference (\ie startup overhead).
The small spike before the region of high utilization marks the warmup period.

We also examine the GPU utilization of two GPUs when we scale out (\cref{fig:laminar-uc:uc1:gpu-util:laminar2}).
We observe that: 1) both GPUs are well utilized though not fully utilized, and 2) the startup overhead increases as we scale out.
For the first observation, our profiling reveals that GPU computation is no longer the bottleneck.
Instead, the data scan from disk emerges as the new bottleneck, struggling to keep up with the rate at which GPUs process data.
The second observation explains the low-performance benefit when we scale to two GPUs due to an increase in startup overhead.

\PP{Limitation}
As the evaluation highlights, the current approach can incur a high startup overhead when activating a large number of workers.
Hiding this overhead is challenging, but one potential solution involves intelligently reducing the level of parallelism when dealing with small datasets.
This estimation can be incorporated during the static query optimization phase.
Furthermore, the current approach focuses on scaling the physical queue resources for the \laminar router but does not address the physical queue resource for the \eddy router on the central queue~\cref{fig:impl:eddy_internal}. 
This could become a contention point as the system scales out further.

\subsection{UC4: Data-Aware Load Balancing}
Lastly, we demonstrate a use case that uses the round-robin routing policy for \laminar when simply scaling up cannot provide an adequate performance benefit due to workload imbalance. We will use the following query.
\begin{lstlisting}[style=SQLStyle, label=lst:laminar-uc:uc4, caption={Query to identify negative food reviews.}]
SELECT * FROM foodreview
 WHERE LLM('What is the following review about? Only choose "food" or "service"', review) = 'food'
 AND rating <= 1;
\end{lstlisting}
In this query, users seek to identify negative customer reviews attributed to poor food quality.
Users can easily find unsatisfying reviews based on the rating associated with each review.
The query uses \lstinline[style=SQLStyle]{LLM} operator built into \eva, utilizing a language model to gain a basic understanding of each review, which is the primary performance bottleneck.
To improve the query performance, \sys uses the \laminar operator to parallelize the \lstinline[style=SQLStyle]{LLM} predicate by scaling up the number of workers responsible for executing the \lstinline[style=SQLStyle]{LLM} predicate on the existing hardware platform.

\PP{Workload Imbalance}
The default round-robin routing policy alternates between workers for scheduling without considering the existing load on the workers or the variation in execution cost due to data differences. 
This may lead to overloading busy workers and underutilizing idle ones.
In the provided example (see ~\cref{lst:laminar-uc:uc4}), two factors contribute to varying execution costs for different data.
Firstly, for language models, the execution cost of the input data is correlated to the length of the input.
For a lengthier review, the language model takes longer to summarize and decide whether it is complaining about the food or the service.
Round-robin routing overlooks data characteristics, such as length, leading to potential overloading or underutilization of workers. 
Secondly, as mentioned previously~\cref{sec:impl:design}, multiple rows of data are grouped into a routing batch to reduce the queuing overhead.
In this query, the simple predicate \lstinline[style=SQLStyle]{rating <= 1} is always executed first due to query optimization rules (lower cost).
This execution order results in some rows being filtered out by the rating predicate in each routing batch, leading to varying workloads for workers.

\begin{figure}[t]
	\centering
	\includegraphics[width=0.85\columnwidth]{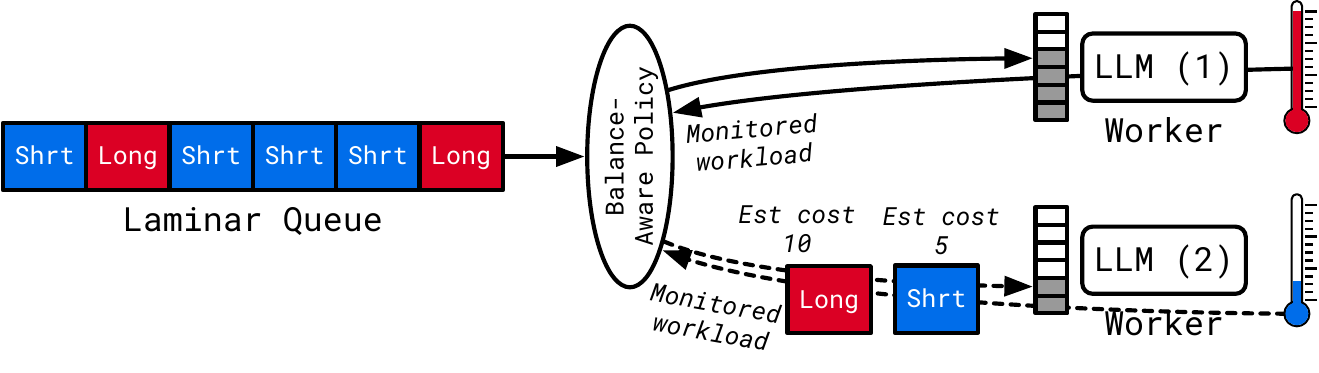}
\caption{\textbf{Workload imbalance of round-robin between workers}}
\label{fig:laminar-uc:uc4:load-imbalance}
\end{figure}

\PP{Data-Aware Load Balancing}
Motivated by the above challenges, we propose implementing a data-aware load balancing in the \laminar routing.
As shown in~\cref{fig:laminar-uc:uc4:load-imbalance}, the router in the \laminar operator will monitor the workload of each worker.
It will always prioritize routing data to the worker with a lower workload.
In this example, as the first worker (\lstinline[style=SQLStyle]{LLM(1)}) is heavily loaded, the router starts to route data to the second worker (\lstinline[style=SQLStyle]{LLM(2)}) until their monitored workloads reach similar levels.

In data-aware load balancing routing, a key challenge is the need for proactive routing decisions rather than reactive. 
Waiting until a workload imbalance is detected among workers to make adjustments would be too late, leading to performance degradation.
To address this, monitored workload metrics for each worker should rely on heuristics rather than solely on profiled statistics.
Users can define custom heuristics for the \udfs; by default, \sys uses input size as a reasonable proxy for execution cost.
For LLMs, this corresponds to the text length, and for vision models, it is the input image/frame size.
As shown in ~\cref{fig:laminar-uc:uc4:load-imbalance}, longer questions are assigned higher estimated execution costs.
Once a question is enqueued, the router adjusts the monitored workload of that worker.

\subsection{UC4: Performance Results}
In this section, we demonstrate the performance benefits of data-aware load balancing. 
Our experiment involves $600$ McDonald's reviews from Google Maps, each accompanied by its published rating.
The \lstinline[style=SQLStyle]{LLM} predicate utilizes Orca, a local large language model with $13$ billion parameters from the GPT4All~\cite{gpt4all_2023} library.
This model runs on CPUs and allows parallelization on multiple threads. 
The user can specify the number of threads in the \udf, so the \laminar operator can also automatically scale the number of workers for this predicate during execution for better performance.
For this experiment, we run the query on the same server with $32$ cores.

\begin{figure}[t]
\begin{subfigure}[t]{0.52\columnwidth}
	\vskip 0pt
  	\centering
	\includegraphics[width=0.9\columnwidth]{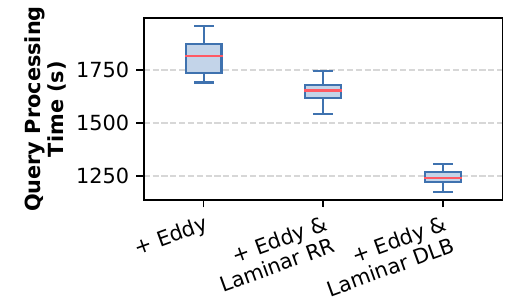}
	\caption{\textbf{Query processing time of different \laminar routing policies} -- comparison between: $+$ \eddy, $+$ \eddy and \laminar with round-robin, and $+$ \eddy and \laminar with data-aware load balancing.}
	\label{fig:laminar-uc:uc4:perf}
\end{subfigure}
\hfill
\begin{subfigure}[t]{0.45\columnwidth}
  	\vskip 0pt
  	\centering
  	\includegraphics[width=0.85\columnwidth]{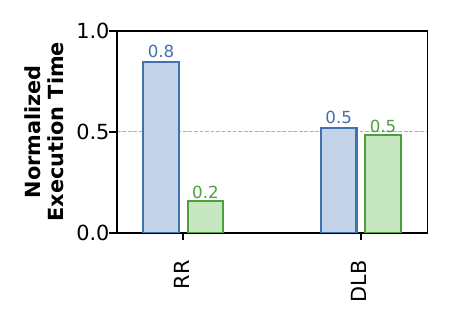}
  	\caption{\textbf{Execution time difference among workers} -- comparison between round-robin and data-aware load balancing.}
  	\label{fig:laminar-uc:uc4:workload}
\end{subfigure}
\end{figure}

To evaluate the performance benefits of data-aware load balancing, we compare three setups: \sys $+$ \eddy, \sys $+$ \eddy and \laminar with default round-robin policy, and \sys $+$ \eddy and \laminar with data-aware load balancing.
The query execution pipeline involves multiple workers and queues, potentially randomizing the data order and influencing query performance.
Thus, we report query processing time on $10$ runs of the same query across the $3$ setups.
As shown in~\cref{fig:laminar-uc:uc4:perf}, the median query processing time without \laminar is $1814.12$ seconds.
Introducing the \laminar operator with the round-robin policy reduces the median query processing time to $1652.67$ seconds because the \laminar operator provides automatic scaling of the \lstinline[style=SQLStyle]{LLM} predicate when hardware resource is underutilized~\footnote{In this case, the \laminar operator only needs to scale to $2$ workers that already saturate the hardware resource.}.
However, due to the uneven workload distribution when the round-robin policy is employed (depicted in~\cref{fig:laminar-uc:uc4:workload}), optimal performance is not achieved.
Using the data-aware load balancing policy further reduces the median query processing time to $1238.98$ seconds ($1.46\times$ improvement).
\section{RELATED WORK}\label{sec:related}
\PP{Visual DBMSs}
In VDBMSs, running deep learning models on every video frame is computationally expensive. 
To address this, researchers have proposed techniques~\cite{noscope_kang_2017, blazeit_kang_2019, tahoma_anderson_2019, DBLP:journals/pvldb/KangGBHZ20, DBLP:journals/pvldb/KangGBHSZ21, DBLP:conf/icde/KoudasLX20, DBLP:conf/sigmod/LaiHLZ0K21, DBLP:journals/pvldb/YangWHLLW22} using lightweight, specialized models to reduce the frames processed by the resource-intensive oracle model.
Focus~\cite{hsieh_focus_2018}, Seiden~\cite{bang_seiden_2023}, and TASTI~\cite{kang_tasti_2022} explore classical techniques like indexing to speed up video analytics queries. 
Other works focus on specific tasks, like object tracking~\cite{miris_bastani_2020}, out-of-domain vocabulary classification~\cite{panorama_zhang_2019}, and action detection~\cite{chunduri_zeus_2022}.
However, these works trade off accuracy for efficiency, while \sys prioritizes system-level optimizations without compromising query accuracy.

Equi-Vocal~\cite{zhang_equi-vocal_2023} presents a new interface through which users can find events in the video corpora by providing positive and negative examples.
Skyscraper~\cite{etl_kossmann_2023} introduces a video extract-transform-load (ETL) problem, focusing on transforming video streams into application-specific formats by applying \udfs during ingestion.
Another line of research focuses on building indexes over precomputed object detections and trajectories to efficiently execute spatio-temporal queries~\cite{DBLP:conf/icde/ChenYK22,DBLP:conf/sigmod/Chen0KY21,DBLP:journals/pvldb/ChenKYY22}.  
However, these works assume that the relevant \udfs are known priori, while \sys does not have this requirement but accelerates any ad-hoc queries.

\PP{Deep Neural Network Serving}
Clipper~\cite{crankshaw_clipper_2017} is a framework that does deep neural networks (DNN) serving equipped with a model abstraction layer.
It applies adaptive and delayed batching techniques to improve the throughput of model serving without violating the latency requirement.
INFaaS~\cite{romero_infaas_2021} instead specializes in improving the model selection logic for the model serving framework.
It automatically determines the model variant and underlying hardware architecture based on performance and accuracy requirements.
Inferline~\cite{crankshaw_inferline_2020} monitors runtime traffic and conducts dynamic scaling during execution to save model serving costs.

Clockwork~\cite{gujarati_serving_2020} attempts to build a system with predictable latency to reduce tail latency.
Scrooge~\cite{hu_scrooge_2021} focuses on optimizing the cost of the deep learning inference by allocating just enough resources for inferencing without violating latency constraints.
Tensorflow serving~\cite{olston_tensorflow-serving_2017} and TensorRT inference server~\cite{nvidia_tensorrt-serving_2020} are production-grade model serving systems.
Nexus~\cite{shen_nexus_2019} proposes squishy bin packing to improve the utilization of GPUs and accurately avoid service-level objective violation.
Cocktail~\cite{gunasekaran_cocktail_2022} proposes to enhance the model selection logic of Clipper and also improves the resource auto-scaling mechanism to ensure requests are handled within SLO.
These systems primarily focus on model serving given the user performance and accuracy requirements.
This aspect is orthogonal to \sys, which specializes in optimizing long-running analytical queries on a database.

\PP{GPU Resource Management and Sharing}
Jain \etal~\cite{jain_dynamic_2018} study existing GPU spatial and temporal sharing mechanisms. 
To maximize spatial sharing, G-Net~\cite{zhang_gnet_2018} proposes offloading network functions to GPU and allows multiple functions to share the GPU resource.
Gandiva~\cite{xiao_gandiva_2018} develops a suspend-and-resume mechanism to allow temporal GPU resource sharing for different DNNs, aiming for quicker feedback on hyper-parameter tuning.
On top of that, Salus~\cite{yu_salus_2019} proposes a new DNNs execution preemption scheme, achieving a fine-grained GPU time-sharing without the need for data migration from GPU to CPU.
Themis~\cite{mahajan_themis_2020} instead focuses on designing a finish-time fair GPU resource-sharing mechanism. 
GSLICE~\cite{dhakal_gslice_2020} proposes GPU resource auto-provisioning during runtime along with NVIDIA MPS technology~\cite{nvidia_mps_2022} to maximize the processing throughput.
Choi \etal~\cite{choi_multigpu-serving_2022} instead study efficiently sharing multi-GPUs resources for model serving.
\sys also adopts the idea of spatial sharing, but it focuses on adapting it to the \aqp framework. 
\section{Conclusion} \label{sec:conclusion}

We presented \sys, an adaptive query processing framework tailored for ML queries.
\sys eliminates the need to collect \udf statistics during query optimization.
Instead, it leverages the \eddy operator to collect statistics during query execution and dynamically routes data to different predicates.
Additionally, \sys takes advantage of the \laminar operator to ensure optimal hardware utilization, scalability, and efficient load balancing among multiple backend workers.
Our empirical results demonstrate that \sys successfully optimizes the query plan across four diverse use cases, achieving a speedup of up to $11.52$\X. 

\clearpage
\bibliographystyle{ACM-Reference-Format}
\bibliography{ref}

\end{document}